\documentstyle{article}
\begin{document}

\noindent{\Large     {\bf
\noindent Double unification, time compression, and space flatness
for the extended particle }}

\bigskip
\noindent  {\large {\bf I.E. Bulyzhenkov } }

\bigskip
\noindent {Candesal Technologies, 473 Dawson Ave.,
Ottawa, ON K1Z 5V6, Canada}

\noindent {E-mail: bulyzhenkov@netscape.net}

\bigskip
\hspace {5 mm} { \parbox {10.9 cm} {
\bigskip
\noindent
{\bf Abstract: } Three variational vector equations are derived for
the extended particle-field object
located on the light cone.  Point sources are excluded from the pure
 field equations and all physical magnitudes are free from divergences.
Accepting 3D intersections of 4D cone-charges, vector electrogravity
explains all observed phenomena under common flat 3D and 1D intervals.
External cone-charges contribute to pseudo-Riemannian metrics of  proper
 four-spaces of charged objects, resulting in dilation-compression of their
proper time rates. Photon-type gravitational radiation is not
 associated with metric modulation of flat, laboratory space. The
 Minkowski-Lorentz equation for free charges corresponds to the equivalence
  principle for the canonical four-space. The predicted criterion of double
   unification, particles with fields - gravity with electrodynamics, is
   confirmed in vector electrogravity.

}}

 \bigskip \bigskip

{\noindent \large {\bf
1.  Introduction}}

\bigskip

Covariant equations for matter were originally derived for independent
carriers of mass and charge [1].
But one elementary object N can carry both
electric, $q_{_N}$, and gravitomechanical (mass), $m_{_N}$,
charges. Gravity or acceleration can lead, for example, to a separation
 of opposite electric  charges
within an electroneutral medium with free electrons [2,3]. The induced
electromagnetic fields under such separation depend essentially
on the mass -  charge  ratio of carriers, while  the mass of a
carrier is not relevant
in Maxwell's equations. The joint carrier for formally separated gravity
 and electromagnetism suggests that it is necessary to search
for new variables for the classical Lagrangian
 of charged matter. The immediate task may be to derive at least one
dynamic equation including the  ratio of electric
and  gravitomechanical   charges of  material carriers.

The canonical   four-momentum  $P_{_N\mu}\equiv m_{_N} u_{\mu}$ $ + $
$q_{_N} { A}^{\neq_N}_\mu$ $\equiv  m_{_N}  {\cal V}_{\mu} +
 m_{_N}U^{\neq _N}_{\mu}$
 seems to be
 one of the most appropriate notions for the description of a charged
 object  N
in its four-space with the proper metric tensor
$g_{\mu\nu}^{_N}(x) $  $ \equiv \eta_{\mu\nu} + g^{\neq _N}_{\mu\nu}(x)$
($g^{_N}_{\mu\nu}(x)\equiv g_{\mu\nu}$, for short;
$ \eta_{\mu\nu} = diag(+1, -1,-1,-1)$),
determined by all external objects K
({\it i.e.}  K $\neq$ N, that is noted by $\neq N$).
When the electromagnetic contribution is absent, $q_{_N} A^{\neq_N}_\mu$ =
0,
the pure  gravitomechanical four-momentum  may be formally separated into
mechanical and gravitational contributions, respectively,
\begin{equation}
   m_{_N}{{u_\mu }}
= m_{_N} \left \{
 {    {  1 + (    {\sqrt { g_{oo} }} - 1 )  }    \over
   \sqrt  {1 - {v_i v^i}}       }  ;
-  {{   v_i +  ({\sqrt {g_{oo}} }  g_i ) }\over
{ \sqrt{1 - {v_iv^i}}}  } \right \}
 \equiv  m_{_N}  {\cal V}_{\mu} +  m_{_N}B^{\neq _N}_{\mu},
\end{equation}
 with the  "curved" three-velocity
 $v_i=\gamma_{ij}v^j$
($u_\mu = V_\mu \equiv g_{\mu\nu} dx^\nu / ds$ for $q_{_N}A_{\mu}^{\neq _N}$
= 0,
with $ds$  =  $(g_{\mu\nu} dx^\mu dx^\nu)^{1/2}$,
$dx^\mu = dx^\mu_{_N}$,
 $v^i\equiv $  $dx^i /g_{oo}^{1/2}(dx^o-g_idx^i)$;
 $ g_i$ $=$ $-g_{oi}/g_{oo}$;  $\gamma_{ij}=g_ig_jg_{oo}-g_{ij}$;
$ \mu, \nu$ $\rightarrow$ $ 0, 1, 2, 3$;
 $i,j$ $ \rightarrow$ $ 1, 2, 3$; $c \equiv 1$).

By its natural involvement in various physical problems,
the canonical four-momentum
may be tried  as a dynamical variable for the
 action of an elementary object N.
But the classical theory of fields and particles, for example [4-8],
does not employ  the canonical four-momentum as a dynamic variable.
The known approaches, for example [7],  to combine mechanical and electric
charges under a  joint geodesic motion
were associated with complicated modifications of space geometry and
discussions about the structure of charges in general relativity.
For well known reasons the classical theories of fields and particles,
 including the non-dualistic approaches [9-12],
 look incomplete and do not overcome some
internal difficulties.

We examine  a non-dualistic way once again by trying to exclude
point sources (associated with observations) from the field equations in
 agreement with Einstein's intention.
The particle integration into the very structure of the field was assumed in
the last Einstein constructions, for example, "We could regard matter as
 being made up of regions
of space in which the field is extremely intense... There would be no room
 in this new
physics for both field and matter, for the field would be the only reality"
 (translation [11]). This program is not accomplished yet in a classical
approach and it may be considered as a motivation for our efforts.

In order to
reveal  the new opportunities of the  classical theory we replace
the point charge by the elementary  charged continuum
emanating from
 a point source in parallel with the Coulomb and the Newton fields.
 This elementary field continuum with homogeneous charge densities at
 the proper light cone-horn points
may be called (conventionally) a cone-particle.
At first glance this  alternative approach  would seem
unreasonable in any practical treatment
because  every infinite  charged continuum  of
matter    would have
infinite energy. But it will be shown below that
the emanating cone-particle and the paired emanating cone-field
together form
a unified material complex, called the elementary
 particle-field object, with zero components of the energy-tensor density
$(T_{_N}^{\mu\nu} = 0$ for superfluid, potential motion).
Einstein's concept  of  cone-charges integrated into "the very
field structure"
becomes free from infinite self-energies. This concept can
propose a  clear mechanism for a particle's  action-at-a-distance  [12],
when extended charges, not sources, interact locally at all
points of the common space-charge-mass 3D continuum.

Again, we start from the assumption  that every elementary charge or
particle
may be considered in terms of an infinite material continuum
(emanating  with a zero four-interval
 from a moving point source).
Each of the two mirror  cones
with the joint vertex in  four-space contains its own particle
matter  to counterbalance
its own elementary field. This assists us in removing from the theory
 the unreasonable
advanced field solutions of classical electrodynamics, where only one
point particle-source (rather than two mirror particle-sources) in a
joint vertex
was wrongly associated with  both Minkowski's cones.

The retarded functions
appear  in the theory
with locally  bound  particle-field matter (at all cone points $x$) only
with
 respect to its  source at cone's vertex $\xi $.
The emitting material cone  object (excluding its vertex)
may be treated as one multifractional field
 in non-dualistic terminology,
but we shall  refer traditionally to the  elementary particle and to the
elementary field  fractions  in order to trace their contributions
to the pure field equations (derived below).
But now cone-particles are not complete elementary objects as they
cannot move independently from their cone fields, located on the same
points of the light cone.

We shall introduce a unified particle-field action
 for one  elementary cone object N,
for which we employ the proper vector variables,  $P_{_N\mu}$  and
$a_{_N\mu}$, associated with external, $m_{_K}a_{_K\mu}$, $e_{_K}
a_{_K\mu}$ and proper, $m_{_N}a_{_N\mu}$, $e_{_N}a_{_N\mu}$, fields,
 respectively. The  Euler-Lagrange equations for the extended cone
 object  will involve only finite physical magnitudes,
and  these four-vector equations will correspond to the known
demands for observed motion of charges.
Electrodynamics and gravitation
appear as a unified non-dualistic field theory, where both  the
Maxwell-type equations and the vector replacement of the Einstein-type
 equation follow directly from  one variational equation, $P_{_N\mu}
 T_{_N}^{\mu\nu} = 0$.

The inseparably bound particle and field fractions of elementary matter
 will assist us in overcoming the classical problem
of charged particle self-acceleration  after  replacing
 the Minkowski equation (with the Lorentz force)
with its generalization,  $m_{_N}DP_{_N\nu}/ ds_{_N}  = P_{_N}^{\mu}
(\nabla_{\mu} P_{_N\nu} - \nabla_{\nu} P_{_N\mu}  )  = 0$, for geodesic
 motion  in the proper canonical four-space $x_{_N}$, where $P_{_N\mu}
 = m_{_N}g^{_N}_{\mu\nu} dx^\nu_{_N}/ds_{_N}$.

After deriving the field equations  in terms
of the proper field densities for every extended  cone object
we shall verify the  symmetrical involvement of external
electric charge and mass densities (associated with a joint forming-up
 field $a_{_K\mu}$)  into the proper canonical four-momentum density
 $P_{_N\mu}$.
This will reveal new (electromagnetic) references for
 the  metric tensor $g^{_N}_{\mu\nu}$ of the proper
pseudo-Riemannian
four-space.

 We shall employ the accepted tetrad formalism in order to demonstrate
  the hidden metric symmetry, $\gamma^{_K}_{ij} = \delta_{ij}$, in the
   three-interval
  $dl^2 = \gamma^{_K}_{ij}dx^i_{_K}dx^j_{_K}$  of any material
object K,
 despite the fact that every component of the proper
pseudo-Riemannian metric tensor, $g^{_K}_{\mu \nu } \neq \eta_{\mu \nu }$,
 represents gravity in full agreement with the Einstein covariant scheme.
This finding will provide the opportunity to
introduce the common (for all objects) space+time manifold $\{{\bf x};
dt\}$, with the universal, flat time interval, $dt^2 = \gamma^{_K}_{oo}
dx^o_{_K}dx^o_{_K}$ and $\gamma^{_K}_{oo} = \delta_{oo}$ for all
 cone-particles K crossing the selected point ${\bf x}$ of the united
  space-charge-mass continuum. The six bounds $ \gamma_{ij}^{_K} =
  g_{oi}g_{oj}g^{-1}_{oo} -g_{ij} = \delta _{ij}$ for ten metric
  components $g^{_K}_{\mu\nu}$ will lead finally to
four, not ten, independent variational equations for vector gravitation.

 It will be remarkable to derive that the covariant applications of
 Euclidean 3D subspace within curved four-space
are consistent with the planet perihelion precession, gravitational
light bending and time dilation.
The flat three-space  acknowledges selection of
the proper mechanical contribution $m_{_N} {\cal V}_{\mu} \equiv
 m_{_N}(1-\delta_{ij}v^iv^j)^{-1/2}\{1; -v_i\} $
and the gravitational contribution
 $m_{_N}  B^{\neq _N}_{\mu} \equiv m_{_N}(1-\delta_{ij}v^iv^j)^{-1/2}
 \{  {\sqrt {g_{oo}} } -  1;
 -{\sqrt {g_{oo}} }  g_i  \}$ in the gravitomechanical momentum (1).

We shall
study the
hitherto unexplained relativistic experiments with rotating superconductors
 [13, 14] in order to  demonstrate the   applications of
 the introduced extended charges for solid state physics.
One could select the other  experimental indications
against the point model of elementary electric charge and mass, including
 the celebrated
Aharonov - Bohm phenomenon [15].

Gravitational interactions in vector electrogravity (VEG) are associated
 with linear synthesis of external vector fields,
 $B^{\neq _N}_\mu = -G\sum_{_K}^{_K\neq_N}m_{_K}a_{_K\mu}$, where every
forming-up four-vector potential  $a_{_K\mu}$ is determined by its
proper field equation. Electrodynamic interactions with external fields
are also associated with the same forming-up potentials, with
$A^{\neq _N}_\mu =  \sum_{_K}^{_K\neq_N}e_{_K}a_{_K\mu}$.
The mass is a source of gravity in the present scheme, rather than the
energy tensor density.

New opportunities of electrogravity with
the extended masses and
flat 3D subspace
allow this theory  to incorporate electrical charges into the
standard covariant scheme with the proper canonical pseudo-Riemannian
four-space.
External mass and electric charge densities cannot change Euclidean
geometry of the proper 3D subspaces, but they affect the proper time
 of every charged object. The predicted electromagnetic time
 dilation-compression
is available for simple laboratory tests. Both electromagnetic and
 gravitational waves in the present scheme are vector photons, contrary
  to the tensor gravitational wave of general relativity, and there are
   no metric modulations of flat, laboratory three-space.

\bigskip\bigskip

\noindent {\large {\bf 2. Action of the extended particle-field object}}

\bigskip

It is common knowledge that
 the covariant electrodynamic equations with a current density
and with the Lorentz force may be obtained from the variational principle
in four-space.
 Both relativistic methods, developed by M. Born [16]  and H. Weyl [17],
 state that it is possible to fix electromagnetic fields under path
  variations for
 charges,
as well as to fix four-coordinates
 of free charges under field variations.
But such assumptions cannot be valid in general. Sometimes a coordinate
displacement
of charges is the only reason for the creation of macroscopic
electromagnetic fields within an electroneutral system (a rotating
 conductor, for example).

The particular purpose of this section is to propose the universal
 dynamic variables in order to eliminate
the preliminary assumptions one uses when
varying the action of charged matter.
For this purpose we consider for a moment the pure particle
   action-at-a-distance
 ${\cal S}_{_N}^{p}$ with one   point particle-source [12],
\begin{eqnarray}{\cal S}_{_N}^{p} = - \int [{\hat m}_{_N}(\xi_{_N}
 [p_{_N}]) u_{\mu}(\xi_{_N}[p_{_N}])+ {\hat q}_{_N} (\xi_{_N}
 [p_{_N}])A_\mu^{\neq _N}(\xi_{_N}
 [p_{_N}])]{{d\xi_{_N}^\mu [p_{_N}]}\over d p_{_N}}  dp_{_N}
 \nonumber \\ \equiv -\int dp_{_N}\left \{{\hat m}_{_N}
u_{\mu}+{\hat q}_{_N}\!\sum^{_K \neq _N}_{_K}\!\int\!dp_{_K}
{\hat q}_{_K}(\xi_{_N})
{\eta}_{_K} (\xi_{_N} , \xi_{_K})_{\xi_{_N} \neq \xi _{_K}}
{  {d\xi _{_K\mu}}\over dp_{_K}}
 \right\}\!{{d\xi_{_N}^\mu}\over dp_{_N}},
\end{eqnarray}
 where the selected source N and  all other sources
K = 1, 2, ..., N-1, N+1, ... are associated, respectively,
with gravitomechanical charge-sources ${\hat m}_{_N}(\xi_{_N}
 [p_{_N}])$ and
$ {\hat m}_{_K}(\xi_{_K}[p_{_K}])$,
electric charge-sources ${\hat q}_{_N}(\xi_{_N} [p_{_N}])$ and
 $ {\hat q}_{_K}(\xi_{_K}[p_{_K}])$,
coordinates $ \xi_{_N}^\mu[p_{_N}]$ and
 $\xi^\mu _{_K} [p_{_K}] $,
 which may  be regarded as functions of parameters
$p_{_N}$ and $ p_{_K}$
on their classical paths $\xi_{_N}  \equiv \xi_{_N} [p_{_N}]$ and
$\xi_{_K}\equiv \xi_{_K}[p_{_K}]$.

The local interaction of a point source N at the point $\xi_{_N}$
  with  the  field
cone-charge density $q_{_K}(\xi_{_N})^{s_{_K}=_0}_{\xi_{_N} \neq \xi_{_K}}
 = q_{_K} = const$
of an external cone object K
is determined by the basic  operator ${ \eta}_{_K}
 (\xi_{_N} [p_{_N}] , \xi_{_K}[p_{_K}])_{\xi_{_N}\neq \xi_{_K}}$.
Any  different point sources
cannot have  all four common coordinates  and  this is
indicated in the symbolic form $\xi_{_N} \neq \xi_{_K}$.
The basic  operator $\eta_{_K}(\xi_{_N} [p_{_N}] ,
\xi_{_K}[p_{_K}])_{\xi_{_N}\neq \xi_{_K}}$
may be  specified on an infinite proper
 four-space $x_{_K}$,
 which intersects, in particular, the point  $\xi_{_N} [p_{_N}] $. We
  accept   continuous
 coordinates $x_{_K}^\nu = \{x_{_K}^o;x_{_K}^i \}$ for every proper
  four-space
with the proper metric tensor $g^{_K}_{\mu\nu} (x_{_K})$.
Intersections of the curved proper four-spaces, due to their
joint 3D subspaces,
  acknowledge an introduction of
the common  space+time, $ \{ dt; {\bf x} \}$, for an ensemble of material
 objects
after an  appropriate application  of the common time interval $dt$
(defined below)
and the common three-space (which ought to keep universal geometry
for all intersecting subspaces).
Below there will appear two opposite parametric time
 intervals, $dt^{_K}_{_1,_2}({\bf x})=\pm |dx_{_K}^o| $,
for mirror three-dimensional motion of any selected object and
antiobject N contrary to the accepted  Minkowski approach
with $dt \equiv dx^o$ for all cases. Moreover,
$dt$ will not be a component of any proper vector $dx_{_K}^\nu$ in spite of
$dt^2 = (dx^o_{_K})^2 $.
This parametric interval $dt$ will be introduced for a considered point
in flat
three-space, where cone-particles (located on different
pseudo-Riemannian four-spaces $x_{_K}$ - light cones, for shot),
 can intersect.

Unlike  the formal interaction-at-a-distance between point sources,
a real interaction of infinite cone objects takes place locally at the
intersection of the emitted  cones at joint material points.
The basic operator ${ \eta}_{_K}
 (x_{_K} , \xi_{_K}[p_{_K}])_{x_{_K}\neq \xi_{_K}}$
will be responsible for the local interaction of the selected object N
with the external object K at joint material points $x_{_N} = x_{_K}$
 under the zero
four-intervals
$s_{_N} (x_{_N},\xi_{_N})=0$ and $s_{_K} (x_{_K}, \xi_{_K}) = 0$,
with $x_{_N}\neq \xi_{_N}$ and $x_{_K}\neq \xi_{_K}$.

By making use of the equality (2) and of the proper four-space
 notion $x_{_N} \equiv x$,  one can
 introduce  for every elementary object N a  covariant four-potential  of
the external  electromagnetic field
$A_\mu^{\neq _N}(x) $
$\equiv$ $A_\mu^{\neq _N}(x,\xi_{_1},\xi_{_2},...,\xi_{_N-{_1}},
\xi_{_N+{_1}},...)$
 $ \equiv$ $\sum^{_K\neq _N}_{_K}  {\cal A}_{_K\mu }
(x)_{x\neq \xi_{_K}
[\tau _{_K}]}^{s_{_K}[\tau _{_K}]={_0}} $,
where $\tau _{_K}$ is a "material" value of the path pa\-ra\-me\-ter
$p_{_K}$ of an elementary
 cone object K when it crosses  the considered point $x$.
The     four-potential
${\cal A}_{_K\mu } (x)_{x\neq \xi_{_K}
[\tau _{_K}]}^{s_{_K}[\tau _{_K}]={_0}}$ of the elementary
 electromagnetic field  of any charged cone object
K at its material point $x_{_K}$, with $x_{_K} = x$, is related to
a four-space
 position of a source K at the  point $\xi_{_K}$
by the zero four-interval, $s_{_K}(x,\xi_{_{_K}}[\tau _{_K}]) $ = 0,
with $x \neq \xi_{_K}$, that determines the "material" parameter
$\tau_{_K}$.

Note, that different   points $x$
correspond to different "material" values of the path parameters
of the same  object, {\it i.e.} $\tau _{_K}\equiv\tau _{_K}(x)$.
One should use the  zero-intervals in determining the elementary
electromagnetic  four-potential
${\cal A}_{_K\mu}(x)_{x\neq \xi_{_K}[\tau _{_K}]}^{s_{_K}[\tau _{_K}]={_0}}
\!\equiv\!\int\!dp_{_K}
 q_{_K} (x)  {\eta}_{_K} (x,\xi_{_K}[p _{_K}])_{x\neq \xi_{_K}[p _{_K}]}
 \{\!d \xi_{_K\mu}[p_{_K} ]/ d p_{_K}\!\}$
or the four\--\-po\-ten\-tial  $a_{_K\mu}
(x)_{x\neq \xi_{_K}[\tau _{_K}]}^{s_{_K}[\tau _{_K}]={_0}}\!\equiv\!\int
 dp_{_K}{ \eta}_{_K} (x,\xi_{_K}[p _{_K}])_{x\neq
\xi_{_K}[p_{_K}]} $
$\{d \xi_{_K\mu}[p_{_K} ]/ d p_{_K}\}$   for the basic (forming-up)
uncharged field  of every  elementary
 object K, which contributes to the total material field
at the considered point $x$.
Only retarded zero-interval relations with sources will appear
for  emitted cone continua after an
appropriate  use   of the two  space+times $\{ dt_{_1,_2},x^i \}$
(one for matter,  the other for antimatter), rather than
 the accepted four-space manifold $\{x^o,x^i\}$.

The proper field (or potential) $a_{_N\mu}
(x)_{x\neq \xi_{_N}[\tau _{_N}]}^{s_{_N}[\tau _{_N}]=o}$ at the
considered point $x$ was emitted by the source at one of its
path points, $\xi_{_K}^\mu[p_{_N}]$, which cannot be defined
without referring to the equation of motion (derived after variations).
This four-vector field takes four degrees of freedom
 and may be a dynamic variable for a
cone object N at all material points $x = x_{_N}$ of its
curved light cone-horn.

A certain source position $\xi_{_N}\equiv \xi$, for short,  may be
conjugated
 (through a zero four-interval) with the material field
points by a defining relation ${\hat Q}_{_N}(\xi)$ $ \equiv \int d^{4}x
Q_{_N}(x)$ ${\hat \delta}_{_N}^{4} (x,\xi)_{x\neq\xi}$.
The accepted pseudo-geometry (zero-interval matter)
defines the structure of the operator
${\hat \delta}_{_N}^4(x,\xi[p])_{x\neq \xi[p]}$ $\equiv $
$\{{\hat \delta}_{_N}^3({\bf x},{\mbox {\boldmath $\xi $} }[p])$
 $ { \delta}_{_N}(x^{_0}-\zeta ^{_0}[p])\}_{x\neq \xi} $,
where the effective coordinate  $\zeta ^{_0}[p]$ should be
determined by the "time"  coordinate  $\xi ^{_0}[p]$  and
by the "time" delay (for  flat  four-space $\zeta ^{_0}[p]$= $\xi^{_0}[p]$
  $\pm |{\bf x}-{\mbox {\boldmath $\xi $} }[p]|$).
By noting  $x\neq \xi$  we emphasize below that  the continuous
 function-densities
represent  emitted cone
matter  at  any considered  point $x$ of four-space but
 not  a source at  the vertex  $\xi$, which is a nonmaterial peculiarity
within this elementary  material continuum.
A similar statement is true for  material cone-particle for points
 ${\bf x}$ and  source points
${\mbox {\boldmath $\xi $} } $
 when it is noted that  ${\bf x}\neq {\mbox {\boldmath $\xi $} }$
for three-space.

One ought to exclude the source point
$\xi$ (and ${\mbox {\boldmath $\xi $} }$)
from an
elementary cone-particle  in order to
avoid a twofold account of the conjugated
notions  (the nonmaterial source and the associated cone-particle) under
 description of one elementary object.
To operate with  two different kinds of coordinates
for  point sources of matter  and for matter itself ({\it i.e.} infinite
 particle-field
cones excluding the vertexes),
 we have to  distinguish  the conjugated characteristics.
For example, a function ${\hat P}_{_N\mu}(\xi_{_N}[p_{_N}])$
represents formally a  canonical four-momentum  of a  point source N.
A function-density $P_{_N\mu}(x)^{s=o}_{x\neq \xi}$
is a  canonical four-momentum density
for a real particle-cone  N at its material points $x\equiv x_{_N}$,
when $x\neq \xi[\tau]$ and   $s(x,\xi[\tau]) = 0 $
 (and for a virtual particle N  at all points $x\neq \xi[p]$ if
$s(x,\xi[p])
  \neq 0 $).

The  gravitomechanical, $m_{_N} (x)^{s={_0}}_{x\neq \xi}$, or the electric,
$q_{_N} (x)^{s={_0}}_{x\neq \xi}$,
 elementary charge  density of the cone-particle
is conjugated to  the point particle-source mass,  ${\hat m}_{_N}(\xi)$, or
electric charge, ${\hat q}_{_N}(\xi)$, respectively,
$\int d^4x m_{_N} (x)
   {\hat \delta}_{_N}^4
(x,\xi)_{x\neq\xi}  $ $\equiv$   ${\hat m}_{_N}(\xi)$ or
$\int d^4x q_{_N} (x)
   {\hat \delta}_{_N}^4 (x,\xi)_{x\neq\xi}$
    $\equiv$   ${\hat q}_{_N}(\xi)$.
The   scalar cone-charge densities  $m_{_N}(x)
\equiv   m_{_N}$ and $q_{_N}(x)
\equiv  q_{_N}$  are  homogeneous functions  of the four-space coordinates,
$ \partial_\mu m_{_N}(x) = 0 $
and  $\partial_\mu q_{_N}(x) = 0 $,
as well as  the charge-source functions $ {\hat m}_{_N}(\xi[p])$
 and ${\hat q}_{_N}(\xi[p])$  are  independent of the path
coordinates $\xi [p]$.  This provides linear relations for the elementary
electromagnetic, ${\cal A}_{_N\mu}
(x)_{x\neq \xi}^{s={_0}}$  $\equiv $ $ q_{_N}a_{_N\mu}
(x)_{x\neq \xi}^{s={_0}}$, and the elementary gravitomechanical,
${\cal B}_{_N\mu}
(x)_{x\neq \xi}^{s={_0}}$  $\equiv $ $ m_{_N}a_{_N\mu}
(x)_{x\neq \xi}^{s={_0}}$, material fields because the densities $q_{_N}$
and $m_{_N}$
are universal constants.

 The    Green's structure of the basic
operator ${ \eta}_{_N} (x, \xi[p])_{x\neq \xi[p]}$
will be described below in the Appendix 1.
What is important to underline right now is that
  the proper canonical four-momentum
density of the cone-particle, $P_{_N\mu}(x)_{x\neq \xi}^{s={o}}$  $\equiv$
$\{m_{_N}(x)u_\mu (x)$ +
$q_{_N}(x)
A_\mu^{\neq{_N}}(x)\}_{x\neq \xi}^{s={o}}$,
is independent of the
forming-up uncharged  cone-field
at every considered
point $x$
of the  elementary particle-field object N.
This means that $P_{_N\mu}(x)_{x\neq
 \xi}^{s={o}}$
 and
 $ a_{_N\mu}(x)_{x\neq \xi}^{s={o}}$
 might be independent dynamic variables
for the description of the same elementary cone object N.

The  particle-source action (2)
is independent of the emitted field fraction of  matter
and can be associated only with the pure
particle fraction of the cone object.
But the particle is always accompanied by its own
field, {\it i.e.} every particle  is only   a fraction of an infinite
   particle-field object.
 Then,
a complete
  action ${\cal S}^{pf}_{_N}$ of this
self-contained object should be contributed to by both the particle
and the  field elementary fractions.

Before adding the paired elementary cone-field to the action (2),
we introduce a  canonical tensor density
$W_{_N\mu\nu} (x)^{s=o}_{x\neq \xi}$  $ \equiv \{ \nabla_\mu
 {P}_{_N\nu}(x) - \nabla_\nu
{P}_{_N\mu}(x) \}^{s=o}_{x\neq \xi}$ = $ \{ \partial_\mu
{P}_{_N\nu}(x) - \partial_\nu
P_{_N\mu}(x) \}^{s=o}_{x\neq \xi}$ of the elementary cone-particle N.
The covariant derivatives, $\nabla_\mu $,
may be  replaced in the vorticity  $W_{_N\mu\nu} (x)^{s=o}_{x\neq \xi}$ by
the partial ones, $\partial_\mu$,
due to the symmetry of the Christoffel coefficients in the proper
four-space $x_{_N}^\mu$ with respect to the proper four-vectors
$P_{_N\mu}$ and $a_{_N\mu}$ (not with respect to $P_{_K\mu}$ and
$a_{_K\mu}$, which
are four-vectors in other curved four-spaces).
The  cone-field contribution to the  complete
 action ${\cal S}^{pf}_{_N}$ of
the elementary particle-field object N can be introduced
in terms of the scalar Lagrangian density
within a  four-dimensional  volume,
\begin {equation}
{\cal S}^{pf}_{_N}\!= -\!\int\!dp {\hat P}_{_N\mu} (\xi[p])
{{d \xi^\mu[p]}\over
d p} - \int {{ {\sqrt {-g_{_N}}}d^4x  }\over 16 \pi }
{{{  \{ g_{_N}^{\mu\rho}g_{_N}^{\nu\lambda}
   W_{_N\mu\nu}(x)f_{_N\rho\lambda}
 (x)\}^{s[\tau ]=o}_{x\neq\xi[\tau ]}
 }} } ,\end {equation}
 where both
the new   tensor density
 $f_{_N\mu\nu}(x)^{s[\tau ]={o}}_{x\neq \xi[\tau ]}\!\equiv\!\{ \nabla_\mu
{a}_{_N\nu}(x)\!-\!\nabla_\nu {a}_{_N\mu}
(x)\}^{s[\tau ]=o}_{x\neq \xi[\tau ]}\!$ =
$\{ \partial_\mu
{a}_{_N\nu}(x)\!-\!\partial_\nu {a}_{_N\mu}
(x)\}^{s[\tau ]=o}_{x\neq \xi[\tau ]}$
 for the elementary cone-field
and the  canonical tensor density
${W}_{_N\mu\nu}(x)^{s[\tau ]=o}_{x\neq \xi[\tau ] }$ for
the elementary cone-particle
are accompanied  by the material restrictions
 $x\neq \xi[\tau]$
and $s(x,\xi[\tau])=0$ for an elementary
cone object N, when it   crosses every point $x$ used in (3) for the
four-dimensional
integration, {\it i.e.} when $x\equiv x_{_N}$. Hereinafter
$g_{_N}^{\mu\nu}(x) \equiv
 \eta^{\mu\nu} +
 g^{\mu\nu}_{\neq _N}(x)$, with
$g_{_N}^{\mu\nu} g_{_N\mu\lambda} = \delta^\nu_\lambda$.

So far, the first item  in the action (3)  corresponds formally to a point
 "charged" source, rather than to a charged cone-particle.
 One  may interchangeably  rewrite the complete action via the operators
for a virtual  object N, which gains its real cone state (crossing the
 considered point $x$) only
after integration in (3) over the path parameter p,
$$
 {\cal S}_{_N}^{pf}=
-  \int dp      \int {\sqrt {-g_{_N}}} d^4x
 \{  {{ P_{_N\mu}(x) }}{{  dx^\mu   }
\over dp}{   {{\hat \delta}_{_N}^4({ x},{\xi}[p])_{x\neq\xi[p]}   }
\over  {\sqrt {-g(x)}} }$$ $$
+  {{ g_{_N}^{\mu\rho}g_{_N}^{\nu\lambda}  W_{_N\mu\nu}(x)
   }\over 16\pi}  \left [{{d\xi{_\lambda }[p] }\over dp  }
{{\partial { \eta}_{_N} (x,\xi[p])}\over \partial x^\rho} -
{{d\xi_{\rho  }[p] }\over dp  }
{{\partial {\eta}_{_N} (x,\xi[p])}\over \partial x^\lambda }
 \right ]_{x\neq\xi[p]} \}   .  \eqno {(3a)}
$$

Different  points $x$ in four-space can be occupied by the same
 material particle - field
cone,   $s(x,\xi[\tau ])$ = 0,  under different
source locations and under different "material" values,
  $\tau \equiv \tau (x) $, of the path parameter $p$.
The  action (3a)  may be formally simplified  in terms of
real cone matter after integration (3a) over $p$,

                            $$
 {\cal S}_{_N}^{pf}\!=\!-\!\int\!{\sqrt {-g_{_N}  }}
d^4x\!\left \{\!P_{_N\mu}(x) {i}_{_N}^\mu
 ({ x})
+  {{ g_{_N}^{\mu\rho}g_{_N}^{\nu\lambda}  W_{_N\mu\nu}(x)f_{_N\rho\lambda}
(x) }\over 16\pi} \right  \}_{x\neq \xi[\tau ] }^{s[\tau ]=o}, \eqno (3b)
$$
where we introduced the  four-flow
density,
\begin{equation}
{ i}_{_N}^\mu(x)_{x\neq \xi[\tau ]}^{s[\tau]={o}}
\equiv
  \int     {{dx^\mu }\over dp}   {{
{{ {\hat \delta}_{_N}^4(x,\xi[p])_{x\neq \xi[p]} }}
  }\over  {\sqrt {-g_{_N}(x) } } }dp,
\end{equation}
of the elementary cone-particle N at any  of its material points
 $x\equiv x_{_N}$,
selected for consideration. But one should keep in mind for formal action
 (3b) that all variations of the action have to
be done with respect to both virtual and material states of the object N,
{\it i. e.} variations have to be considered before the integration over
the parameter p. After the integration over p the independent variables,
 like $P_{_N\mu}$ and $x^\mu$, may be bound by additional restrictions
 for material states.

Accepting approach with action for one selected object N, we ought
to distinguish contributions from the proper, $a_{_N\mu}$,  and
external, $a_{_K\mu}$, fields.
The proper four-momentum  $P_{_N\mu }(x)$
  and the proper metric tensor
  $g_{\mu \nu } = g^{_N}_{\mu \nu }(x)$ are independent from $a_{_N\mu}$,
   but
they  both
depend on the same system of external  objects K with their proper fields
$a_{_K\mu}(x)$. That is the reason that
 the proper canonical momentum and the
proper metric tensor are related variables (their links with the same
external field $U^{\neq_N}_\mu $ = $\sum_{_K}^{_K\neq_N}c_{_K} a_{_K\mu}$
 will be revealed at the final steps of the below developed scheme for
 canonical four-space, where $P_{_N\mu}P_{_N\nu} g^{_N}_{\mu\nu} =
 m_{_N}^2$). One may assume for a moment that the external field
  $U^{\neq_N}_\mu $  is to
be a final variable for the object N, rather than $P_N\mu$ or
$g^{-N}_{\mu\nu}$. But the sum $U^{\neq_N}_\mu $, as well as each summand
$a_{_K\mu} $, is not a four-vector in the proper space $x^\mu_{_N}$
($a_{_K\mu}\neq g_{_N\mu\nu}a_{_K}^\nu$ when K $\neq$, for example).
The proper four-vector $P_{_N\mu}$ can perfectly represent the external
Newton-Coulomb fields in the action (3)-(3b) because we shall find that
 $\delta P_{_N\mu}$ = $\delta U^{\neq_N}_\mu $ and there will be no
 tensor fields in the present approach to gravitation.
 Finally only proper coordinates, $x^\mu_{_N}$, and two proper
 four-vectors, $a_{_N\mu}(x)$ and $P_{_N\mu}$, may be selected
  as independent variables in the developed theory with the vector
   variational equations.

One may propose to study an ensemble of elementary objects and to consider a
summary action $\sum_{_N}{\cal S}^{pf}_{_N}$. But there is no universal
four-space for all elementary objects and, contrary to classical
electrodynamics, there is no way to introduce
a collective field variable in the present scheme.
It is more accurate to operate with a system of
 the  Euler - Lagrange equations, derived below
  for every elementary object in its external fields,
rather than to speculate about a joint action and possible collective
 variables for an ensemble of different objects.

\bigskip \bigskip

\noindent {\large {\bf 3. Universal time interval }}

\bigskip

The  variational procedure in (3a) with respect to the canonical
four-mo\-men\-tum density,
  both real and virtual variations ${\delta} P_{_N\mu }(x) $, with
  $\delta g_{_N}^{\mu\nu} = 0$,
can lead to a Maxwell-type equation for a total four-flow density,
${ I}_{_N}^\nu(x)_{x\neq
\xi[\tau]}^{s[\tau]={_0}}$, of the elementary cone object,
\begin{eqnarray}
  { I}_{_N}^\nu(x)_{x\neq
\xi[\tau]}^{s[\tau]={_0}} \equiv
 \left \{  { i}_{_N}^\nu(x)
- {{ \nabla_\mu
[ f_{_N}^{\mu\nu}(x)-
f_{_N}^{\nu\mu}(x) ]} \over
 8\pi } \right \}_{x\neq \xi[\tau]}^{s[\tau ]={_0}} = 0,
\end{eqnarray}
where
$f_{_N}^{\mu\nu}(x)_{x\neq \xi}^{s={_0}}
\!\equiv\!g_{_N}^{\mu\rho}(x)
g_{_N}^{\nu\lambda}(x)$
${f}_{_N\rho\lambda}(x)^{s={_0}}_{x\neq \xi}$
= $-f_{_N}^{\nu\mu}(x)_{x\neq \xi}^{s={_0}}
$.
Note that not all components of the skew-symmetric
tensors are independent under variation [11]:
 the relations $\delta W_{_N\mu\nu}(x)$ =
$ - \delta W_{_N\nu\mu}(x)$
must be taken into account.

So far we do not know the final relations of $P_{_N\mu}$
and $g_{_N}^{\mu\nu}$ with external fields, and we temporary
neglect the  relation between $\delta P^{_N}_\mu$ and
$\delta g_{_N}^{\mu\nu}$. For this reason the Maxwell-type
  equation (5) is incomplete in our approach (this simplified
   equation will be accepted only for  potential states, when
    $W_{_N \mu\nu}(x)_{x\neq \xi}^{s={_0}} = 0$).

The arbitrary  variations are not necessarily
compatible [18] with
any restricting conditions for the path parameter $p$,
for example $s[p]\neq 0 $ and $P_{_N\mu}P^\mu_{_N}\neq m^2_{_N}$
under virtual variations in (3a).
But after variation of the action, one may specify the appropriate
path parameter in the derived equations of motion due to some additional
 restrictions
for real matter  (or for real antimatter).  In equation (5) we operate with
the family of material points $x$ which correspond to the real
cone  object N. A selection of any one point $x$
provides an appropriate selection of the  path parameter $p=\tau_{_1,_2}$
due to the material restriction $s[\tau_{_1,_2} ]=0$ with two possible
solutions
$\tau_{_1}$ and $\tau_{_2}$ for the mirror cones in the metric four-space.

Even though the covariant equations are four-dimensional in the proper
four-space, dynamics
of matter depends on the development parameter, and there must be a
 three-dimen\-sio\-nal
picture as seen by an observer. This motivates us to introduce
a new parametric interval $dt$  in order to describe
 the evolution of matter (or antimatter)  in three-space ${\bf x}$.
 One can therefore perform the integration  over  $p$ in the
definition (4) and to introduce  the material  four-flow densities of
 two mirror
cone-particles
via the appropriate time differentials $dt_{_1}$ and $dt_{_2}$,
\begin {eqnarray}
{ i}^\nu_{_N}(x)^{s[\tau _{_1,_2}]=
{_0}}_{x\neq \xi[\tau _{_1,_2}]} \equiv
 \int  {{dx^\nu }\over dp{}}
 {{ \{   {\hat \delta}_{_N}^3
({\bf x},{\mbox {\boldmath $\xi $}}[p]) {\delta}_{_N}(x^{_0}-
\zeta ^{_0}[p] )\}_{x\neq \xi[p]}
      }\over  {\sqrt {-g}}  }
dp \nonumber \\
=
\int     {  dx^\nu \over  {\sqrt {-g}}   dp  }
  {{   \{ {\hat \delta}_{_N }^3 ({\bf x},
{\mbox {\boldmath $\xi $}}[p])
\delta_{_N } (p-\tau _{_1,_2})
\}_{x\neq \xi[p]}
}
\over   \left |  {{ \partial \zeta ^{_0}[p ]
 }\over \partial p{}} \right |_{x\neq   \xi [p] }      }
dp
= \nonumber \\
\left \{ {{ {\hat \delta}_{_N  {_1},{_2}}^3 ({ \bf x},
{\mbox {\boldmath $\xi $}}[\tau_{_1,_2}]) dx^\nu
 }\over   {  \sqrt {\gamma}}    {\sqrt {g_{oo}}}    d\tau _{_1,_2}
 \left |  { { {\partial  \zeta ^{_0}}[\tau _{_1,_2}]
 }\over \partial \tau _{_1,_2}} \right |
 } \right \}^{s[\tau_{_1,_2}]=_0}_{{\bf x}\neq {\mbox {\boldmath $\xi $}}
[\tau _{_1,_2}]} \equiv
\left \{  {{  {\hat \delta}_{_N{_1},{_2}}^3
({\bf x}, {\mbox {\boldmath $\xi $} })  }
\over  {\sqrt \gamma } }
 {{ d{ x}^\nu }\over {\sqrt {g_{oo}}}  dt_{_1,_2}  }
 \right \}^{s[\tau _{_1,_2}]=_0 }_{{\bf x}\neq
 {\mbox {\boldmath $\xi $}}[\tau_{_1,_2}]},
\end{eqnarray}
where ${\gamma}\equiv ||\gamma_{ij}|| = {{-g}}/g_{oo}$.
The  operators ${\hat \delta}_{_N  {_1},{_2}}^3 ({\bf x},{\mbox
 {\boldmath $\xi $}}[\tau_{_1,_2}])^{s
[\tau_{_1,_2}]=_0 }_{{\bf x}\neq {\mbox {\boldmath $\xi $}}[\tau_{_1,_2}]
 } $    conjugate function-paths of the point sources  in three-space
(${\mbox {\boldmath $\xi$}} [\tau _{_1,_2}]$ $\equiv$
${\mbox {\boldmath $\xi$}}_{_N  {_1},{_2}} [\tau _{_1,_2}]$
for the mirror particle-sources $N_{_1} $ and $N_{_2}$)
to  function-densities
of the mirror  cone-particles  N$_{_1}$ and N$_{_2}$ projected onto
three-space  ${\bf x}$.

  Every considered  point ${\bf x}$  with  coordinates  $x^i$ in
the 3D subspace of the proper four-space
$x^\nu_{_N}\equiv x^\nu$ can be related to two
source  points $\xi[\tau_{_1,_2}]$,   $\xi[\tau_{_1}]$  $\neq$
  $\xi[\tau_{_2}]$,
by zero-interval conditions,
${s{}(x,\xi [\tau_{_1}])=0}$
and  ${s(x,\xi[\tau_{_2}])=0}$.
Both these conditions  provide
$d x^o=d \zeta ^o[\tau _{_1,_2}]$, where
$d\zeta ^o[\tau _{_1,_2}]$ $=$ $d\tau _{_1,_2}\partial \zeta ^o
[\tau _{_1,_2}]/ \partial \tau _{_1,_2} $ for the
  mirror cones with matter or antimatter.
By making use of this fact we introduced in (6) the opposite parametric
differentials at every point ${\bf x}$,
\begin{equation}
dt_{_1,_2}({\bf x}) \equiv
\left \{{{ d\tau _{_1,_2}}\over |d\tau _{_1,_2}|    }
\left | d\tau _{_1,_2}
  {{ \partial \zeta ^{_0}[\tau  _{_1,_2}]
 }\over \partial \tau _{_1,_2}} \right |\right \}^{s
[\tau_{_1,_2}]=_0 }_{{\bf x}\neq {\mbox {\boldmath $\xi $}}[\tau_{_1,_2}]
 }
= \{  sign (d\tau _{_1,_2})  |dx^o|  \}^{s
[\tau_{_1,_2}]=_0 }_{{\bf x}\neq {\mbox {\boldmath $\xi $}}[\tau_{_1,_2}]
 },
\end{equation}
which may be called the direct and
 the inverted  time intervals, $dt_{_1}({\bf x})$ and $dt_{_2}({\bf x})$,
  respectively.
The parametric intervals $dt_{_1,_2}({\bf x}, \tau_{_1,_2})$ are introduced
 for the cone-particles (cone-charges) crossing a particular point in
 three-space ${\bf x}$. It is important for the anticipated  description
  of an ensemble
that both these intervals are finally independent from
the proper parameters of different objects and the universal time
interval may be employed for all cone-particles
with rest masses.

The direct, with $\tau _{_1}$, (the inverted, with $\tau _{_2}$)
 four-flow  density (6)  of a cone-particle and the direct
(the inverted) time interval (7) are associated with the direct
(the inverted)  elementary cone-field in (5) (Appendix 1).
By applying three-space and the time parameter from (7),
one finds a coincidence  of a dynamic three-dimensional picture
 for  direct particle-field
objects (matter) and for inverted ones (antimatter)
under appropriate applications of the direct
and the inverted time intervals,   $dt_{_1}$ = $-dt_{_2}$ = $|dx^o|$, for
example.
 The appearance
of two opposite time intervals (7)
with parametrically oriented directions
  provides an  opportunity
to  introduce two parallel space+time manifolds, $\{ dt_{_1}, x^i\}$
and $\{dt_{_2}, x^i\}$, on the basis of one four-space metric system
$\{ x^o, x^i\}$.
This allows one to trace the bound charge-time contribution into
 Charge-Parity-Time symmetry and to explain the PT symmetry
violation.

The mirror  elementary
cone-particles  N$_{_1}$ and N$_{_2}$ occupy  the direct
and  the inverted cones with matter and antimatter, respectively, within
one metric four-space.
But a particle (or antiparticle) fraction from
one cone is not bound with an antifield  (or field) fraction
from the mirror cone. By trying to relate one
 point charged particle
in the joint vertex of
two  pure field cones to both these fields, the Minkowski theory resulted in
the unreasonably
 advanced solutions for emitted field matter.

There are neither retarded nor advanced  relations
of the cone-particle with the paired cone-field
in the  concept of extended  cone-charges. The
cone-field and  cone-particle elementary densities are locally bound
(without any delay) at every
material point in  four-space. By choosing appropriately  the
space+time  manifolds  for matter or   for antimatter,
  one obtains
only  retarded emission  from  point sources.
Below we   omit
the $"1"$ or $"2"$ subscript in $d\tau $ or $dt$  by dealing, for
simplicity,
only with matter in the direct space+time manifold $\{dt, x^i\}$, $dt
 = + |dx^o|$,
for example.

 Again, the choice of the  time parameter
$t$ for matter or antimatter is irrelevant. The important point is
that the  cone-particle four-flow density  may be divided
in (5) into the
direct  and the inverted components, as well as  the  elementary
cone-field at the left hand side of the equation (5).
Note, that the Dirac operator $\delta_{_N}^4(x-\xi)\equiv
\delta_{_N}^3({\bf x}-
{\mbox {\boldmath$\xi $}})  $ $\delta_{_N} (x^o-\xi^o)$  for
one  point object at $x=\xi $
 cannot
provide the splitting of the four-flow density  (6) into the two mirror
 components,
contrary to the operator ${\hat \delta}_{_N }^4(x,\xi)
_{x\neq\xi}$ for mirror  cone-particles.
 It is in principle impossible to consider two  mirror
 point charges
 in   one reference point $\xi $
because the mirror particles carry opposite charges, $m_{_N  {_1}}=
-m_{_N {_2}}$
and $q_{_N  {_1}}= -q_{_N  {_2}}$.

Varying (3a) with respect to ${\delta} P_{_N\nu }(x)$, with $\delta
 g^{\mu\nu}_{_N} = 0$, one may
reintroduce the Maxwell-type equation (5) in the following operator
form,
\begin {eqnarray}
\nabla_\mu
  g_{_N}^{\mu\rho} g_{_N}^{\nu\lambda}\!\left [
{{d\xi_\lambda [p]  }\over dp} {{\partial
{\eta_{_N}} }
\over \partial x^\rho} - {{ d \xi_\rho [p]  }\over dp}
{{\partial { \eta_{_N}}}
\over \partial x^\lambda} \right ]_{x\neq \xi[p]}
   =  4\pi
   {{dx^\nu }\over dp}
 {{ {\hat\delta }_{_N}^4 (x,\xi[p])_{x\neq \xi[p] }}\over
 {\sqrt{-g_{_N}}} }  .
\end {eqnarray}

The Euler-Lagrange equation (5) or (8) suggests a way to
speculate  about the structure of the  vector basic cone-field
  $a_{_N\mu}
(x)^{s={_0}}_{x\neq \xi}$  or
 the     scalar basic operator
 ${\eta}_{_N} (x, \xi[p])_{x\neq \xi}$ in any
curved four-space.
Both these equations can be
easily  solved  in    flat  four-space
via  Green's function, ${G}(x,x')_{x\neq x'}$$=$$\{\delta
(x^o-{x'}^o\mp|{\bf x}- {\bf x}'|) /
|{\bf x}- {\bf x}'| \}_{x\neq x'}$,
associated with the fundamental operator equation $\partial_\mu \partial^\mu
{ G}(x,x')_{x\neq x'}$ =
$4\pi {\hat\delta}^4 (x,x')_{x\neq x'} $
$\equiv 4\pi \{ {\hat \delta}^3 ({\bf x},{\bf x}')\delta
(x^o-{x'}^o\mp|{\bf x}-{\bf x}'|) \}_{x\neq x'}$
(Appendix 1).

Every considered  point  ${\bf x}$ with three-coordinates
$x^i$ can be related   to sources
 of different material cones by zero-interval conditions.
In other words, different  material cone objects can cross one common point
 ${\bf x} \equiv {\bf x}_{_1},{\bf x}_{_2}..., {\bf x}_{_N}, ...$
like light or gravity of distant stars cross the Earth at any fixed time.
Superposition of different elementary cone objects in one common
three-dimensional
space ${\bf x}$ may be described under the
common  time interval $dt$, because all one dimensional intervals
(7), $dt^2_{_K} \equiv \gamma^{_K}_{oo}dx^o_{_K}dx^o_{_K}$, are flat
($\gamma^{_K}_{oo} = \delta_{oo}$ is
independent from values  of the  individual path  parameters
$\tau_{_K}$).
We shall prove below that all proper three-spaces $x_{_K}^i$,
associated with different objects K, have
the universal metric tensor, $\gamma^{_K}_{ij} = \delta_{ij}$, and flat
intervals,   $dl_{_K}^2 = \gamma^{_K}_{ij}dx^i_{_K}dx^j_{_K}$, contrary
to proper four-spaces $x^{\mu}_{_K}= \{x_{_K}^o;  x^i_{_K}\} $ with
different metric
tensors $g^{_K}_{\mu\nu}\neq \eta_{\mu\nu}$ and curves intervals $ds^2_{_K}
= g_{\mu\nu}^{_K}dx^\mu_{_K}dx^\nu_{_K}$.

For this reason only both
the common three-space ${\bf x}$, with flat three-interval, and the common
flat interval $dt({\bf x})$ are appropriate to apply to all  objects,
rather than proper four-spaces $x^\mu_{_K}$
 (unspecified for the ensemble).
Due to the common space+time existence,  one may sum (5)
over an ensemble of different potential states in
 $\{ {\bf x}, dt({\bf x}) \}$
 and  find
 the following equations for the total
four-flow density ${ i}^\nu(x)$ of all intersecting at x cone-particles
\begin {equation}
   { i}^\nu({\bf x}, t) \equiv
 {{ { n}_o({\bf x})} {{{}}}} {{d{ x}^\nu}\over
d t} \equiv
 \sum_{_N}     { i}_{_N}^\nu
(x)^{s_{_N}[\tau _{_N}]={_0}}_{x\neq \xi_{_N}[\tau _{_N}]}
 =  {{ \sum_{_N} \nabla_\mu
   f_{_N}^{\mu\nu} (x)^{s_{_N}[\tau _{_N}]= {_0}}_{x\neq
\xi_{_N}[\tau _{_N}] } }
 \over 4\pi },
\end {equation}
for the four-current density of their gravitomechanical cone-charges
(masses)
\begin {eqnarray}   { j}_m^\nu({\bf x}, t)
\equiv {{   {  \mu}_o({\bf x})} {{{}}} }
 {{d{ x}^\nu}\over d t}
 \equiv \sum_{_N} m_{_N}
 { i}_{_N}^\nu(x)^{s_{_N}[\tau _{_N}]= {_0}}_{x\neq \xi_{_N}[\tau _{_N}]}
\nonumber \\
 = {{ \sum_{_N}\nabla_\mu
m_{_N}f_{_N}^{\mu\nu} (x)^{s_{_N}[\tau _{_N}]={_0}}_{x\neq \xi_{_N}
[\tau _{_N}]}
}\over 4\pi }
\equiv {{ \sum_{_N}\nabla_\mu
{\cal M}_{_N}^{\mu\nu} (x)^{s_{_N}[\tau _{_N}]={_0}}_{x\neq \xi_{_N}
[\tau _{_N}]}
}\over 4\pi },
\end {eqnarray} and for the  four-current density of
 their electric cone-charges
\begin{eqnarray}
{ j}_q^\nu({\bf x}, t)
\equiv
{{{ \rho}_o({\bf x})}  {{{}}}}
{{d{x}^\nu}\over d t}
\equiv \sum_{_N} q_{_N}{  i}_{_N}^\nu
(x)^{s_{_N}[\tau _{_N}]={_0}}_{x\neq \xi_{_N}[\tau _{_N}]}
 \nonumber \\
= {{ \sum_{_N}\nabla_\mu  q_{_N}f_{_N}^{\mu\nu}
 (x)^{s_{_N}[\tau _{_N}]={_0}}_{x\neq \xi_{_N}[\tau _{_N}]}
}\over 4\pi }
 \equiv {{
 \sum_{_N} \nabla_\mu {\cal F}_{_N}^{\mu\nu}
 (x)^{s_{_N}[\tau _{_N}]={_0}}_{x\neq \xi_{_N}[\tau _{_N}]}
}\over 4\pi }
\equiv
{{{\nabla_\mu  F^{\mu\nu}({\bf x}, t)}}\over 4\pi}.
\end{eqnarray}

The three-space functions
 ${ n}_o({\bf x}) \equiv \sum_{_N} \{  {\hat \delta}^3
({\bf x}, {\mbox {\boldmath $\xi $}  }_{_N}
[\tau _{_N}])/
{\sqrt {-g}}\}^{s_{_N}
= {_0}}_{{\bf x}\neq {\mbox {\boldmath $\xi $}  }_{_N} } $,
 ${ \mu}_o({\bf x})$ $ \equiv$ $ \sum_{_N}
  \{ m_{_N} {\hat \delta}^3
 ({\bf x},{\mbox {\boldmath $\xi $}  }_{_N}[\tau _{_N}])/{\sqrt {-g}}
 \}^{s_{_N}={_0}}_{{\bf x}\neq
 {\mbox {\boldmath $\xi $}  }_{_N}} $,
  ${ \rho}_o({\bf x}) \equiv \sum_{_N} \{  q_{_N}
{\hat \delta}^3({\bf x},{\mbox {\boldmath $\xi $}  }_{_N}[\tau _{_N}])/
{\sqrt {-g}}
\}^{s_{_N}={_0}}_{{\bf x}\neq
{\mbox {\boldmath $\xi $}  }_{_N}} $
were introduced  in (9)-(11) in order to represent the particle matter
 density, the
gravitomechanical charge density,
 and the electric charge density, respectively,
 in three-space ${\bf x}$ for an ensemble
of material cone objects. One will recall, that (9)-(11)
were derived for partial kind of motion of each
elementary object K, when $W_{_K\mu\nu} (x_{_K}) = 0$.

Note that (11) coincides formally with the Maxwell-Lorentz equation
 for the  electric
current density.
But the  equations (9)-(11) were obtained for infinite
cone-particles and cone-fields in
  space+time,   rather than for  point  particle-sources and cone-fields.
 The physical densities ${ n}_o({\bf x})$, ${ \mu}_o({\bf x})$,
 and ${ \rho}_o({\bf x})$, for example,  are associated with  field cone
 objects
 rather than
with point objects. In other words
the  non-dualistic equation (11), for example,
relates the  continuous  field density
${ j}_{q}^\nu ({\bf x}, t)$ of cone-charges with
the  field density $\nabla_\mu F^{\mu \nu }({\bf x}, t)$ at every local
point of the space+time manifold.

In turn, the   electric current density
$j_{q}^\nu ({\bf x}, t)$,  which is specified
for cone-charges excluding the source peculiarities, might be formally
   conjugated to
  a sum of moving point
charge-sources, ${\hat q}_{_N}$,  distributed over these  peculiarities.
But contrary
 to the density of the material continuum,
the density of  the point  sources
 at one fixed point $x$ is meaningless.
One should not neglect this obvious fact by trying to formulate
 a self-consistent theory in the classical way of point carriers
of electric charge or mass.
One may operate at least with a finite number of peculiarities
 within
a finite volume rather than within a single point.

The requirement for a finite physical magnitude
 at  all space points, for all material objects additionally  motivate
 us  to take into consideration an elementary electric charge (and mass)
 in  terms of an elementary continuum (cone)
 with one  reference point (source in cone's vertex) and with a homogeneous
  charge
density $q_{_N}$ (and $m_{_N}$) at all points of this elementary continuum.
It seems very unlikely that it is possible to overcome the problem of
divergence in classical electrodynamics
 without changing the accepted paradigm of point charges
(but not   point sources for the extended charges).
"A coherent field theory," stated Einstein  (translation [11]),
 "requires that all its elements
be continuous ... And from this requirement arises the fact that the
material
 particle
has no place as a basic concept in a field theory. Thus, even apart from the
fact
that it does not include gravitation, Maxwell's theory cannot be considered
a complete
theory." The above introduced separation of the extended material particle
and its point source ({\it i.e.} the transformation of the classical point
particle into the charged cone continuum) is simply a
probe way to complete the theory and to include gravitation into
electrodynamics.

An independent  equation for the elementary cone-field in its
curved  proper four-space $x = x_{_N}$,

\begin {equation}
\left \{ \partial_\mu {f_{_N}}_{\nu\delta}(x)
+\partial_\nu {f_{_N}}_{\delta\mu}(x)
+ \partial_\delta {f_{_N}}_{\mu\nu}(x)
\right \}^{s[\tau]={_0}}_{x\neq \xi[\tau]}
 = 0,
\end {equation}
follows directly from the definition of the elementary tensor
${f_{_N}}_{\mu\nu}(x)^{s[\tau]
={_0}}_{x\neq \xi[\tau]}$.

The similar  Maxwell-type equation in common space+time,
\begin {equation} \partial_\mu {F}_{\nu\delta}({\bf x}, t)
+\partial_\nu {F}_{\delta\mu}({\bf x}, t)
+ \partial_\delta {F}_{\mu\nu}({\bf x}, t)
 = 0, \end {equation}
may be considered for the total field tensor
 $F_{\mu\nu}({\bf x}, t) \equiv \sum_{_K}
q_{_K}f_{_K\mu\nu}(x)^{s_{_K}[\tau_{_K}]={_0}}_{x\neq
 \xi_{_K}[\tau_{_K}]}$.

Notice that the dynamical equation (5) is independent
 of $m_{_N}$ and $q_{_N}$. The elementary uncharged
    field with the forming-up  four-potential
 $a_{_N\mu} (x)^{s
={_0}}_{x\neq \xi}$      in this equation
 may be used as a unified basis
 for the generation of both the gravitational (Newton)
and the electromagnetic (Coulomb) cone-fields.

\bigskip \bigskip

\noindent  {\large{\bf 4. General motion and potential states}}

 \bigskip
\noindent{{ 4.1.  Wave equations}}

\bigskip

One could formally divide the  canonical four-momentum
density  $P_{_N\mu\nu}
(x)^{s={_0}}_{x\neq \xi}$
of the elementary particle into a gravitomechanical
 part and an electrical one.
Then the  canonical
 tensor $W_{_N\mu\nu}(x)^{s={_0}}_{x\neq \xi{}}$
 would be formally divided
into  a gra\-vi\-to\-mechanical part (with $m_{_N}$) and
an electric part (with  $q_{_N}$),
\begin {equation}  W_{_N\mu\nu} (x)^{s
={_0}}_{x\neq \xi} = \{ m_{_N}
M_{\mu\nu}(x) +
q_{_N}  F^{\neq _N}_{\mu\nu}(x)\}^{s
={_0}}_{x\neq \xi},
\end {equation} where
$M_{\mu\nu}(x)$
 $\equiv
 [ \partial_\mu {u}_{\nu} (x)
- \partial_\nu { u}_{\mu}(x)]$
= ${w}_{\mu\nu}  + G_{\mu\nu}^{\neq _N} $,
${w}_{\mu\nu} \equiv [ \partial_\mu {\cal V}_{\nu} (x)
- \partial_\nu {\cal V }_{\mu}(x)]$,
${\cal V}_\nu(x) \equiv \{\beta^{-1}, -\beta^{-1}v_i \}$, $\beta
= {\sqrt {1 - v_i v^i}}$,
$G^{\neq _N}_{\mu\nu} \equiv [ \partial_\mu B^{\neq _N}_{\nu} (x)
- \partial_\nu B^{\neq _N}_{\mu}(x)]$,
and  $F^{\neq _N}_{\mu\nu}(x) $ $
  \equiv $ $[\partial_\mu A^{\neq _N}_\nu (x)$
- $\partial_\nu A^{\neq _N}_\mu (x)] $
$ =
\sum_{_K}^{_K\neq _N} q_{_K}$ $ [\partial_\mu {a_{_K}}_\nu (x)
 - \partial_\nu {a_{_K}}_\mu(x)]^{s_{_K}={_0}}_{x\neq \xi_{_K}}$, and
 $x = x_{_N}$.
By replacing here the partial derivatives with the covariant ones,
we used symmetry of Christoffel coefficients exclusively for
the proper four-vector $P_{_N\mu}$ in curved four-space $x^\mu_{_N}$.
 External fields $a_{_K\mu}$ are
not four-vectors in the proper four-space $x^\mu_{_N}$, and  $\nabla_\mu
 A^{\neq _N}_\nu (x)
- \nabla_\nu A^{\neq _N}_\mu (x)$ $ \neq $ $\partial_\mu A^{\neq _N}_\nu (x)
- \partial_\nu A^{\neq _N}_\mu (x)$, for example. For this reason
the gravitomechanical and electric parts of the tensor (14) can not be
considered separately as tensors.

The action (3b) for real object may be varied
 with respect to its proper covariant field
under arbitrary material or virtual variations
${\delta} a_{_N\nu }(x)$.
This provides  a wave  Euler-Lagrange  equation for
general motion of cone objects in external fields,
 \begin{equation}
\nabla_\mu
 W_{_N}^{\mu\nu} (x)^{s_{_N}
[\tau_{_N}]={_0}}_{x\neq \xi_{_N}[\tau_{_N}]}   =
 0,
\end{equation}
where
$W_{_N}^{\mu\nu}(x) \equiv g_{_N}^{\mu\rho}(x)
g_{_N}^{\nu\lambda}(x){W}_{_N\rho\lambda}(x)$ =
 $-W_{_N}^{\nu\mu}(x) $ and $x = x_{_N}$.

The  other  independent equation for the canonical
 tensor density follows from its
 definition,
\begin {equation}
\{\partial_\mu {W_{_N}}_{\nu\delta}(x)
+\partial_\nu {W_{_N}}_{\delta\mu}(x)
+ \partial_\delta {W_{_N}}_{\mu\nu}(x)\}^{s={_0}}_{x\neq \xi}
 = 0
\end {equation}  and, when $q_{_N} F_{\mu\nu}^{\neq _N}(x) = 0$,
\begin {equation}
\{ m_{_N} [\partial_\mu {M}_{\nu\delta}(x)
+
\partial_\nu {M}_{\delta\mu}(x)
+ \partial_\delta {M}_{\mu\nu}(x)]\}^{s={_0}}_{x\neq\xi} = 0.
\end {equation}

Now we consider the wave equations (15)-(16) for the canonical
tensor density (14) in more detail.
The components of the density $F^{\neq _N}_{\mu\nu}(x) $ can be
associated with  three-vector fields,
 electric
 ${{ E}_{i}(x) }\equiv F_{oi}^{\neq _N}(x)$
 and magnetic ${B}_{e}^i(x)$ $\equiv -e^{ijk}
F_{jk}^{\neq _N}(x)/2{\sqrt \gamma }$ ($e^{123}=1$) ones,
 acting on a cone-particle N
with the homogeneous electric charge density $ q_{_N}(x) = q_{_N}$.
The components of the introduced set of sixteen values
$w_{\mu\nu}$ may be associated with the following three-vectors
\begin{equation}
\cases {
  g_{i}(x)\equiv  \partial_o {\cal V}_i -\partial_i {\cal V}_o \equiv
-{\partial_o} v_i \beta^{-1}
- \partial _i  \beta^{-1}   \cr
{ h}^i({ x}) \equiv
 -{{e^{ijk}}\over 2{\sqrt \gamma }} (\partial_j{\cal V}_k-\partial_k
{\cal V}_j)
\equiv \{ curl \beta^{-1}{\bf v} \}^i. \cr }
\end{equation}
Below we shall proof that ${\sqrt \gamma}$ and $\beta$ are independent
 from metrics and (18) holds only
 "flat" components, which are responsible for pure mechanical
motion. Gravitational fields are implemented into the proper tensor (14)
 through the external potentials $B^{\neq _N}_\mu$.
According to (1) only these potentials are related to changes of the
metric tensor
of electrically neutral objects, for which external three-vector fields
 looks as follows,
\begin{equation}
\cases  {
 G_i(x)\equiv G^{\neq _N}_{oi}(x) \equiv \partial_o B^{\neq _N}_i -
 \partial_i B^{\neq _N}_o =
-{\partial_o}
 \beta^{-1} {\sqrt {g_{oo}}}  g_i
- \partial _i
   \beta^{-1}( {\sqrt { g_{oo}}} - 1 )  \cr
{ H}^i({ x})\equiv-{{e^{ijk}}\over 2{\sqrt \gamma }   }
G^{\neq _N}_{jk}(x)\equiv-{{e^{ijk}}
\over 2{\sqrt \gamma }} (\partial_jB^{\neq _N}_k-\partial_k B^{\neq _N}_j)
= \{ curl \ \beta^{-1}
 {\sqrt{g_{oo}}} {\bf g}
 \}^i. \cr }
\end{equation}

Now the equation (17) for pure gravitomechanical
systems may be read in a three-vector form,
\begin {equation}
\{  m_{_N} div\ ({ \bf h  } + {\bf H}) \}^{s=o}_{x\neq \xi }
 = 0
\end {equation}
 and
\begin {equation}
\{ m_{_N} [ \{ curl\ ({\bf g} + {\bf G}) \}^i
 + \partial_o (h^i + H^i)] \}^{s=o}_{x\neq \xi }  = 0,
\end {equation}
because of equalities $div\ curl \ {\bf a} = 0$ and
$curl\ grad \ {\bf a}
 = 0$ for $\{ curl \ {\bf a}\}^i$ $\equiv$ $(2 {\sqrt {\gamma}})^{-1}$
 $e^{ijk}(\partial _j a_k - \partial _k a_j),$
$div \ {\bf a}$ $\equiv$ $\gamma^{-1/2}\partial_i ({\sqrt \gamma} a^i)$.

One can also represent the general equation (16)
at real field points, $x\neq \xi[\tau]$ and
 $s(x,\xi[\tau])=0$, of electrically charged objects in a three-vector form,

\begin{equation}
div [ m_{_N} {\bf h} + m_{_N} {\bf H} + q_{_N}
 {\bf B} ]^{s=o}_{x\neq \xi }  = 0,
\end{equation}

\begin{equation}
\{( curl [ m_{_N} {\bf g} + m_{_N} {\bf G}
+ q_{_N} {\bf E}])^i +
\partial_o [ m_{_N} { h}^i + m_{_N} H^i +
q_{_N}{ B}^i ]\}^{s=o}_{x\neq \xi }  = 0.
\end{equation}
The difference between (20)-(21) and (22)-(23) is
related with the different
proper spaces $x = x_{_N}$ for charged and uncharged masses. (Below we shall
find that pseudo-Riemannian metrics of charged objects is associated
with external electromagnetic potentials, and one should use  in (19)
replacements $G_i \rightarrow G_i + (q_{_N}/m_{_N})E_i $, $H^i
\rightarrow H^i +
(q_{_N}/m_{_N})B^i $ for relations
with the similar components of the novel metric tensor).

 Contrary to  classical theory, which admits
  bulk (free of particles) three-space regions,
the equation (9), for example, can
not be applied with  zero  density of particle matter at
any point ${\bf x}$. Charged cone matter of the same elementary
 particle-field  object
is emitted from different positions of its source and this elementary
matter
crosses all different  three-space points ${\bf x}$ at the same time
parameter.
The elementary cone-particle (and cone-charge) density exists
simultaneously at all three-space points (the same is true for the
elementary cone-field
density).
For these reasons, a total superposition  of cone-particles
(and cone-charges)
 always has to be  present at any three-space point,
{\it i.e.} $n_{o}({\bf x})\neq 0$  and $\mu _{o}({\bf x})\neq 0$
for all ${\bf x}$   (while $\rho_{o}({\bf x})$ could be equal to
 zero at some
three-space points only due to the opposite signs of the electric
 charge densities in the material superposition). Three-space is
 actually a space-charge-mass continuum without bulk regions, and
  source peculiarities
are not included in this material medium.

\bigskip \bigskip
\noindent { { 4.2. Superfluid or potential states}}

\bigskip

To apply the derived equations to practical
 problems of condensed matter, for example,
we consider the simplest partial solutions of
 (15) and (16) in the absence of the field $P_{_N\mu}$ vorticity,
 $W_{_N\mu \nu }(x)^{s={_0}}_{x\neq \xi} \equiv 0$,
when $I^\mu_{_N}(x)^{s={_0}}_{x\neq \xi} \equiv 0$ due to (5).
 Such potential state, $\partial_\mu P_{_N\nu} = \partial_\nu P_{_N\mu}$,
 is well known for macroscopic superfluid
systems. The  canonical four-momentum density
 $P_{_N\mu}(x)^{s={_0}}_{x\neq \xi}$ can be written in this case
via a scalar  potential
$ \Upsilon_{_N}(x)^{s={_0}}_{x\neq \xi}$, {\it i.e.}
\begin {equation}
P_{_N\mu}(x)^{s={_0}}_{x\neq \xi}\equiv
\{m_{_N}(x)u_\mu(x) +
q_{_N}(x)A_\mu^{\neq _N}(x)\}^{s={_0}}_{x\neq \xi}
 = -\partial_\mu \Upsilon_{_N}(x)^{s={_0}}_{x\neq \xi},
\end {equation}
with $\partial_\mu \partial_\nu \Upsilon_{_N}(x)^{s={_0}}_{x\neq \xi} =
\partial_\nu \partial_\mu \Upsilon_{_N}(x)^{s={_0}}_{x\neq \xi}$.
The potential state, $W_{_N\mu \nu }(x)^{s[\tau]=
{_0}}_{x\neq\xi[\tau]}=0$, corresponds to superfluid
matter without radiation or energy exchange and reads at all points
 $x$ of the elementary cone
N in terms of  the three-vector
functions
                     \begin{equation}
\{ m_{_N} {g}_{i}(x) + m_{_N} {G}_{i}(x)+ q_{_N}
{ E}_{i}(x)\}^{s[\tau]=
{_0}}_{x\neq\xi[\tau]} = 0,
\end{equation}
\begin{equation}
\{ m_{_N}{h}^i(x) + m_{_N} H^i(x) +
+ q_{_N}B^i(x)\}^{s[\tau]=
{_0}}_{x\neq\xi[\tau]} = 0.
 \end{equation}

These dynamic equations describe the mutual counterbalance
of gravitomechanical and electromagnetic forces acting on
the cone-charge densities
    $m_{_N}$  and $q_{_N}$  in the presence of
 external electromagnetic field and gravity.
 The equations (25)-(26) are not new for dissipationless macroscopic
 systems.  Actually  (25)
 is a relativistic generalization of the Bernoulli
stationary equation for a charged ideal fluid, while (26) exhibits
the known fact that the  London stationary
supercurrent is proportional to an electromagnetic three-vector
potential in a superconductor.

The potential states, which are associated with equations (24)-(26),
may be applied  to many known problems,
 such as the field description of the Cooper pairs  within a superconductor,
the  dissipationless  motion of  free electrons  within a conductor,
the  bound electron states in atoms, etc. Each of these elementary carriers,
which are not
involved in collisions,
can be characterized by the individual potentials
$\Upsilon_{_K}(x)_{x\neq \xi}^{s=_0}$ (corresponding to the phases
of wavefunctions for elementary matter  in quantum theory).
The potential state of the extended charged object, when
$P_{_N\mu}(x)^{s={_0}}_{x\neq \xi}= -\partial_\mu
\Upsilon_{_N}(x)^{s={_0}}_{x\neq \xi}$ at all  points
of the proper four-space $x = x_{_N}$, may be considered as a
superfluid state of this cone-object in the common space+time
 $\{ {\bf x},dt \}$.

Now we shall consider  a uniformly
rotating conductor (the Faraday disk) in order to analyze the
relativistic experiment of Ref. [14]  in terms of the approach,
developed herein.
Rotation   leads to
the gravito-inertial fields  (18) and (19).
 The induced  stationary electric and magnetic fields
within the conductor create compensating Lorentz forces
 that allow free charges
 to  rotate synchronously with the lattice charges.
Only a small (conducting)  fraction of electrons on the Fermi
surface deviates from potential states due to
 inelastic collisions.
One may say that  stationary  charged  cone-particles (electrons
within the Fermi volume) take
potential states and satisfy (24) - (26).

We find the electric and magnetic fields
within a uniformly rotating conductor with an angular frequency
$\mbox{\boldmath$\omega$}$
 in an inertial frame
(where $<{\bf v}_{_K}>_{_K}   =   \mbox{\boldmath$\omega$}
\times {\bf r }$, $G_i \rightarrow 0, H^i\rightarrow 0$) by
 averaging (25) and (26),
 respectively, over the
ensemble,

\begin{eqnarray}
< {\bf E}_q(x) >_{_K}
=
<\mbox{\boldmath$\partial$}
{{m_{_K}}\over  q_{_K}  {   \sqrt {1 -  {\bf v}_{_K}^2 }  }   }>_{_K}
\approx  - {{m_o  \omega^2 {\bf r}} \over    |q_o|
({1-    <v _{_K}^2>_{_K}})^{3/2}
       }
,\end{eqnarray}

\begin{equation}
 <{\bf B}(x)>_{_K}
= - < {  {m_{_K} }\over q_{_K}}
 curl \
   {{ {\bf v}_{_K}   }\over   {   \sqrt {1 -  {\bf v}_{_K}^2 }  }    }
         >_{_K} \approx
{ { 2 m_o   \mbox{\boldmath$\omega$} }
\over |q_o|{\sqrt {1-    <v _{_K}^2>_{_K}}    } },
\end{equation}
where $q_{_K} = - |q_o| < 0$ is the negative electron charge,
$m_{_K} = m_o$ is the rest electron mass,
 and
 $
m_o(1 -  <v^2_{_K} >_{_K} )^{- 1/2}$
is the relativistic electron mass averaged over
the Fermi volume ($0 \leq      v^2_{_K}  \leq v^2_{_F} \approx 10^{-4} $,
where $v_{_F}$ is the Fermi velocity).

The electric and magnetic fields within
a uniformly rotating superconductor can also be determined from
 (27) and (28), respectively, by averaging
 over  all free electrons of the total Fermi volume and Cooper
pairs:
both  normal and paired electrons are in potential states.
 Relatively small fractions  of paired and conducting electrons
(both fractions on the Fermi
surface) may provide, in our consideration, only  relatively
 small contributions
to the relativistic corrections in (28).

By using  the relativistic  accurate data of  the experiment [14]
for the magnetic flux within rotating niobium
 superconductors (for which $(1 - v^2_{_F})^{- 1/2} $ = 1.000180
 due to the  independent Fermi surface data),
we may conclude for the mass-charge ratio from (28) for this experiment
that $1.000084(21) m_o/q_o$ = $  m/|q|$ =  $m_o /(1 -
<v^2_{_K}>_{_K})^{1/2} |q_o| $. This result, {\it i.e.}
 $<v^2_{_K}>_{_K} $= 0.47($\pm 12$)$v_{_F}^2$,
confirms the above statement about a dominant contribution to the
 London magnetic moment from dissipationless electron states under
 the Fermi surface,
rather than only from the Cooper pairs on the Fermi surface
(when one might expect $<v^2_{_K}>_{_K} = v^2_{_F}$).

Conventional theory predicts that superfluid fractions of
free electrons has to take only Fermi velocities, $v_{_K} = v_{_F}$, that
evidently disagrees with the experimental results.
The experiment [14] demonstrated, that the potential states of charged
 fermions  within the Fermi volume is
also a superfluid form of matter.
But only superfluid Cooper bosons
 are responsible for the observed quantization of
the  London magnetic moment due to the non-vanishing  macroscopic
potential $ <\Upsilon_{_K}^s >_{_K}$ $ = \Upsilon^s_{_K}$ $ \neq$
$ 0$.
Each fermion possesses a stationary superfluid state
with  an individual potential $\Upsilon_{_K}^n $. However,
any  ensemble of fermions cannot  exhibit macroscopic superfluid
properties in space+time because  $ <\Upsilon^n_{_K}>_{_K} =  0 $
after averaging over the Fermi volume, rather than over the equipotential
 Fermi surface in the case of the paired electrons.
  Nevertheless the Barnet magnetic moment of rotating
 normal metals is also associated with superfluid states of fermions
without the moment quantization on the macroscopic level.

Again, the developed approach to free electrons in potential states
 can explain the measured relativistic
mass-charge ratio of superfluid carriers that is hitherto unexplained
 by the currently available theories.

Superfluid states of charges, when $W_{_N\mu \nu }(x)^{s={_0}}_{x\neq
 \xi} \equiv 0$ and  $I^\mu_{_N}(x)^{s={_0}}_{x\neq \xi} \equiv 0$
 (and $T^{\mu\nu}_{_N} = 0,$ introduced below),
are very special cases of matter motion without energy exchange and
 radiation.
Below we consider more general forms of motion for which the
incomplete equation (5) cannot be applied.

\bigskip \bigskip

\noindent
{\large { \bf 5. Proper energy-tensor of the extended object}}

\bigskip

The Hilbert variation  procedure [19] for (3a)
 with respect to
 variation of the proper  metric tensor, ${\delta}  g_{\mu \nu }(x)
\equiv { \delta}  g^{_N}_{\mu \nu }(x) $
should provide   the symmetric energy-tensor density,
$T^{\mu\nu}_{_N}(x)^{s=o}_{x\neq \xi}$,
 of the elementary particle-field
 object N.
One may fix  under this variation  the contravariant coordinate
 vectors $ dx^\mu $ (but not the covariant ones,
$\delta dx_\nu =  \delta ( g_{\mu\nu} dx^\mu) $ = $dx^\mu \delta
g_{\mu\nu} $), the universal scalars $m_{_N}(x)$
 and $q_{_N}(x)$,
the  covariant four-vector potential
$a_{_N\nu} (x)^{s={_0}}_{x\neq \xi}$ and
 the covariant field tensor
$f_{_N\mu\nu}(x)^{s={_0}}_{x\neq \xi}$.
Note that symmetric components of $g_{\mu \nu }$
are not independent  one from another,
$\delta g_{\mu \nu }$ = $\delta g_{\nu \mu }$,
and we define  $\delta S \equiv - \int dx^4{\sqrt {-g}}
(T^{\mu\nu} \delta g_{\mu\nu}
+ T^{\nu\mu} \delta g_{\nu\mu})/2
= - \int dx^4 {\sqrt {-g}} T^{\mu\nu} \delta g_{\mu\nu} $.

The proper canonical four-momentum $P_{_N\mu}$
depends on external gravitational and electromagnetic
fields (the scheme with
$P_{_N\mu} = m_{_N}g^{_N}_{\mu\nu}dx^\mu / ds$ in the proper
canonical four-space
 will be clarified below).
 Its variations are not independent from the
 variations of the metric tensor,
 ${ i}^\mu \delta P_{\mu } $=$m_{_N}{ i}^\mu\delta
[ g_{\mu\nu}dx^\nu (g_{\rho\lambda}dx^\rho dx^\lambda)^{-1/2} ]$
= $ [(\delta g_{\mu\nu}) m_{_N}{ i}^\mu  dx^\nu/2ds] $
+ $ [(\delta g_{\nu\mu}) m_{_N}{ i}^\nu  dx^\mu/2ds] $
= $m_{_N}({ i}^\mu dx^\nu  + { i}^\nu dx^\mu )
\delta g_{\mu \nu }/2ds$. The
 term
$-{\sqrt {-g}}
( f^{\mu \nu }_{_N}\delta W_{_N\mu\nu}+
 f^{\nu \mu }_{_N}\delta W_{_N\nu\mu}) / 16\pi$
 may be transformed under the four-space integral into
   $ (m_{_N} {\sqrt{-g}}\nabla_\nu
 {f}_{_N}^{\nu\mu}
 /4\pi) \delta  P_{_N\mu }$.
     The contravariant metric tensor is related to the covariant one,
{\it i.e.} $\delta g^{\alpha \beta }
= - g^{\alpha \mu }g^{\beta \nu}\delta
g_{\mu \nu }$ $-$ $g^{\alpha \nu }g^{\beta \mu }\delta g_{\nu \mu }   $
$= - \delta g_{\mu \nu }(g^{\alpha \mu }g^{\beta \nu } + g^{\alpha \nu  }
g^{\beta \mu })$,
 $\delta {\sqrt {-g}} =  {\sqrt {-g}}( g^{\mu\nu}
\delta g_{\mu\nu} + g^{\nu \mu }\delta g_{\nu \mu }) / 2$
= $ {\sqrt {-g}}( g^{\mu\nu}
+ g^{\nu \mu })\delta g_{\mu \nu } / 2  $.

There are no special reasons in our approach
 to involve artificially a scalar metric curvature,
$R_{Ricci}$, into the complete action (3a) for the
collisionless particle-field object.
The curvature ought to appear naturally in any
self-contained theory.
 Moreover, the Rainich - Misner criterion, $R_{{_R}{_M}}$=0,
for   unified theories [20,21] dismisses scalar curvatures in the
initial dynamical equations.

Finally, after  variation (3a) with respect to
  ${ \delta} g^{_N}_{\mu\nu}$ (and  $ \delta
 g^{_N}_{\nu\mu}$), one can obtain
the proper energy-tensor density,
\begin {eqnarray}
 T_{_N}^{\mu\nu}(x)_{x\neq \xi[\tau ]}^{s[\tau ]=0}
\equiv  \{
 {{m_{_N} }\over 2} \left (
{dx^{\mu}\over ds_{_N}}  { I}_{_N}^{\nu }(x)
+ {dx^{\nu}\over ds_{_N}}  { I}_{_N}^{\mu }(x)  \right )
\nonumber \\
+ {  W_{_N\rho\lambda}(x)
\over 16 \pi}
[ g_{_N}^{\mu\nu} {f}_{_N}^{\rho\lambda}(x) -
2g_{_N}^{\mu\rho}{f}_{_N}^{\nu\lambda}(x)
- 2g_{_N}^{\nu\rho}{f}_{_N}^{\mu\lambda}(x)]
 \}_{x\neq \xi[\tau ]}^{s[\tau ]=0},
\end {eqnarray}
for one elementary particle-field object N at its material
 points  $x_{_N} = x$,  with $x\neq \xi[\tau]$ and $s
(x, \xi[\tau])=0$. This tensor takes only zero components
for the above considered potential motion, when
$I_{_N}^\mu = W_{_N\mu\nu} = 0$.
One may say that the paired in the selected object cone-particle
and cone-field fractions energetically
 compensate (or screen) each other for these dissipationless
 states of matter.

If ten different components of the proper metric tensor
$g^{_N}_{\mu\nu}$ can be independent one from another,
as is generally accepted, one could expect
to derive from the action (3a) the Einstein-type tensor equation,
\begin {equation}
 T_{_N}^{\mu\nu}(x)_{x\neq \xi[\tau ]}^{s[\tau ]=0} = 0,
\end {equation}
 for every considered object.

Below we shall derive a novel four-vector equation
for arbitrary motion of the particle-field object,
 $T_{_N}^{\mu\nu }P_{_N\mu}$=0, rather than the Einstein-type tensor
  equation for gravitation, $T_{_N}^{\mu\nu}=0$, which takes
  place in vector electrogravity only for potential, superfluid states.
By deriving the ten-component Einstein equation from the
variational principle, one accepts the   conventional assumption
of general relativity
that components of the symmetric tensor
$g_{\mu\nu}$ may have ten degrees of freedom
(without referring to any physical notions behind this basic
mathematical statement). As a result ten components of the
energy-tensor, rather then masses with Newton's vector fields,
 are origin of tensor gravity in general relativity.
In the last section
we shall find that the metric tensor or its tetrad can take only
four degrees of freedom associated with external Coulomb-Newton
four-potentials $a_{_K\mu}$. A reasonable four-vector equation
for gravitation has to replace ten tensor relations in the most
 general  case of motion.

Total density of matter  at any selected point of the common
three-space ${\bf x}$  takes contribution from
different extended cone-objects. In order to describe the
energy-tensor density under intersection
of traceless superfluid objects at one particular point of
space+time and derive the Einstein-type equation,
one could consider a  sum of the elementary densities,

\begin{equation}
T^{\mu\nu} ({\bf x}, t) \equiv \sum_{_N} T_{_N}^{\mu\nu}
(x_{_N})^{s_{_N}[\tau_{_N}]={_0}}_{x_{_N}\neq
\xi_{_N}[\tau_{_N}]}= \sum_{_N}0 = 0.
\end{equation}
But there are no 4D intersections of different curved four-spaces $x_{_N}$,
 only
their 3D  subspaces with universal geometry can intersect.
A transaction in (31) or in (9)-(11) from different proper
four-spaces $x_{_N}$ to the common space+time manifold
$\{{\bf x}; dt ({\bf x})\}$
 is not a trivial procedure (that will be clarified in the next section).

Assuming for a moment that it is possible to introduce
the curved four-space with common (or averaged) metric tensor,
${\tilde g}_{\mu\nu} $, with $d{\tilde s}^2 =
{\tilde g}_{\mu\nu}d{\tilde x}^\mu d{\tilde x}^\nu$,
one could formally rewrite (31)
in the Einstein-like form,
\begin {eqnarray}
{1\over \kappa} \left (
 {\tilde R}^{\mu\nu} - {{\tilde g}^{\mu\nu}\over 2}
  {\tilde g}_{\rho\lambda}{\tilde R}^{\rho\lambda}
   \right )
=   {{{ \mu}_o( {\tilde{\bf x}})} { {{} }  }   }
 {{ d{\tilde x}^\mu}\over  d{\tilde x}^o } {{ d{{\tilde x}}^\nu}
 \over  d{\tilde s} }
\nonumber \\
 +   {{\tilde F}_{\rho\lambda} \over 16 \pi}
[ {\tilde g}^{\mu\nu} {\tilde F}^{\rho\lambda} -
2 {\tilde g}^{\mu\rho}{\tilde F}^{\nu\lambda}
 - 2 {\tilde g}^{\nu\rho}{\tilde F}^{\mu\lambda} ],
\end {eqnarray}
where the following definitions are introduced,
${\tilde R}^{\mu\nu} \equiv {\tilde G}^{\mu\nu} - 2^{-1}{\tilde g}^{\mu\nu}
{\tilde g}_{\rho\lambda}{\tilde G}^{\rho\lambda}, $
$$
{\tilde G}^{\mu\nu}({\tilde x}) \equiv {\tilde G}^{\mu\nu} ({\tilde{\bf x}},
{\tilde x}^o) \equiv {{\kappa }\over 8\pi}\!\sum_{_N} \{
m_{_N}{{d x_{_N}^\mu}\over  d{ s_{_N}}}
  \nabla_\rho  {f}_{_N}^{\rho \nu}(x_{_N})
+ m_{_N}{{d x_{_N}^\nu}\over  d{ s_{_N}}}\nabla_\rho  {f}_{_N}^{\rho \mu}
(x_{_N})$$
$$
+ m_{_N}M_{\rho\lambda}(x_{_N})[ g_{_N}^{\mu\rho}{f}_{_N}^{\nu\lambda}
(x_{_N})
 + g_{_N}^{\nu\rho}{f}_{_N}^{\mu\lambda}(x_{_N})
 - {{g_{_N}^{\mu\nu}}\over 2}{f}_{_N}^{\rho
\lambda}(x_{_N})]\}^{s_{_N}[\tau_{_N}]={_0}}_{x_{_N}\neq \xi_{_N}
[\tau_{_N}]},
$$
$$   {{{ \mu}_o({\tilde {\bf x}})} { {{} }} }
 {{ d{\tilde x}^\mu}\over  d{\tilde x}^o } {{ d{\tilde x}^\nu}\over
   d{\tilde s} }
\equiv
\sum_{_N}{m_{_N}\over 2}
 \left( { i}_{_N}^\mu(x_{_N}) {{dx_{_N}^\nu}\over  d { s_{_N}}}
+ {i}_{_N}^{\nu }(x_{_N})
 {{dx_{_N}^\mu}\over  d{ s_{_N}}}  \right)^{s_{_N}[\tau_{_N}]=
 {_0}}_{x_{_N}\neq
 \xi_{_N}[\tau_{_N}]},$$
$$   {{\tilde F}_{\rho\lambda}}
[ {\tilde g}^{\mu\nu} {\tilde F}^{\rho\lambda} -
2 {\tilde g}_{_N}^{\mu\rho}{\tilde F}^{\nu\lambda}
 - 2 {\tilde g}^{\nu\rho}{\tilde F}_{_N}^{\mu\lambda} ]$$
 $$ \equiv  \sum_{_N}
  { F^{\neq _N}_{\rho\lambda}(x_{_N})  }
[ g_{_N}^{\mu\nu} {\cal F}_{_N}^{\rho\lambda}(x_{_N}) -
2 g_{_N}^{\mu\rho}{\cal F}_{_N}^{\nu\lambda}(x_{_N})
 - 2 g_{_N}^{\nu\rho}{\cal F}_{_N}^{\mu\lambda}(x_{_N}) ]^{s_{_N}[\tau_{_N}]
={_0}}_{x_{_N}\neq \xi_{_N}[\tau_{_N}]}. $$

A trace of the Einstein-type equation  (32),
 $-{\kappa}^{-1} {\tilde R} = -{\kappa}^{-1}{\tilde g}_{\rho\lambda}
 {\tilde R}^{\rho\lambda}$, with
\begin {equation}
-{ {\tilde R} \over \kappa}\!=\!{ \mu}_o({\tilde{\bf x}})
{{d {\tilde s}}\over {{{}  }} d{\tilde x}^o  }
\equiv\!\sum_{_N}m_{_N}\!\int dp_{_N} {{
 {{ {\sqrt {g^{_N}_{\mu\nu}dx_{_N}^\mu dx_{_N}^\nu } }   }\over dp_{_N}}
  {{ {\hat \delta}_{_N}^4 (x_{_N},\xi_{_N}[p_{_N}])_{x_{_N}\neq \xi_{_N}} }
     \over
 {\sqrt {-g_{_N} } } } } },
\end {equation}
depends on the "curvature"
${\tilde R} ({\tilde x})$ ($k = 8 \pi G$ is the Einstein  constant [8])
 and formally
corresponds to the equality $\sum_{_N} g^{_N}_{\mu\nu}
T^{\mu\nu}_{_N}(x_{_N})^{s_{_N}[\tau_{_N}]
={_0}}_{x\neq \xi_{_N}[\tau_{_N}]}\equiv 0 $.

The above separation of gravitomechanical and electromagnetic fields
in (32) looks artificial for the introduced cone object with
joint forming-up field for electric charge and mass densities.
 It is more reasonable
 to divide, if ever, (31)
into the pure cone-particle and pure cone-field contributions as follows,

\begin {eqnarray}
{1\over \kappa}
 {\tilde {\cal R}}^{\mu\nu}({\tilde{\bf x}}, {\tilde x}^o)
=   {{{ \mu}_o({\tilde{\bf x}})} { {{} }  }   }
 {{ d{\tilde x}^\mu}\over  d{\tilde x}^o } {{ d{\tilde x}^\nu}\over
  d{\tilde s} },
\end {eqnarray}
where
\begin {eqnarray}
{\tilde {\cal R}}^{\mu\nu}({\tilde x})\!\equiv\!{{\kappa }\over
 8\pi}\!\sum_{_N}\{
 m_{_N}{{d x_{_N}^\mu}\over  d{ s_{_N}}}
  \nabla_\rho  {f}_{_N}^{\rho \nu}(x_{_N})
+ m_{_N}{{ d x_{_N}^\nu}\over  d{ s_{_N}}}\nabla_\rho  {f}_{_N}^{\rho
\mu}(x_{_N})
\nonumber \\
+  W_{_N\rho\lambda}(x_{_N})\![ g_{_N}^{\mu\rho}f_{_N}^{\nu\lambda}
(x_{_N}) + g_{_N}^{\nu\rho}f_{_N}^{\mu\lambda}
(x_{_N}) - {{g_{_N}^{\mu\nu}}\over 2}f_{_N}^{\rho
\lambda}(x_{_N})] \}^{s_{_N}[\tau_{_N}]={_0}}_{x_{_N}\neq \xi_{_N}
[\tau_{_N}]}.
\end {eqnarray}
Again, the equation (34) looks artificial for extended masses
under the "averaged" geometry of the joint curved four-space ${\tilde x}$
and the joint curved three-space ${\tilde {\bf x}}$, where
intersection of cone-particles can be expected.

The external forming-up fields ${a_{_K\nu }}(x)^{s={_0}}_{x\neq\xi}$
can be traced  under formation of the field densities in (32) and (35),
which may be formally associated with the Ricci curvature in the Einstein
equation.
One may assume
that the  metric tensor  could also depend on contributions of external
fields with $a_{_K\mu}$
and $a_{_K\nu}$.  But these forming-up fields are common for both
elementary Newton-like,
 ${\cal B}_{_N\nu }(x)^{s={_0}}_{x\neq \xi}
  $  $\equiv$    $ m_{_N}
{a_{_N\nu }}(x)^{s={_0}}_{x\neq\xi}  $,
and Coulomb-like, ${\cal A}_{_N\nu }(x)^{s={_0}}_{x\neq \xi}
  $  $\equiv$    $ q_{_N}
{a_{_N\nu }}(x)^{s={_0}}_{x\neq\xi}  $, four-potentials.
These particular findings can motivate us to
reconsider the metric properties of the proper four-space $x_{_N}$, that
  may   open the  gates for
modernization of gravity within Einstein's covariant
formalism. Unification for gravitation and electromagnetism
may take place, for example, under the canonical four-space with
electromagnetic and gravitomechanical connections.
Below we shall study the symmetrical involvements
of external masses and charges into the proper metric tensor
and develop the new covariant approach to
 the extended  cone-particle within the
 flat 3D continuum of space-charge-mass.

\bigskip \bigskip

\noindent {\large{\bf 6. Gravitation under flat three-space}}

\bigskip

The covariant constructions for every selected object depend essentially on
proper four-space geometry associated with external matter.
External matter (and proper four-space geometry) differs for different
objects.
Curved four-space in general relativity is associated with
curved three-space due to  Schwarzschild's solution  for point masses [22].

There are no point masses in our consideration and Schwarzschild's
solutions cannot be appropriate for extended cone-particles. Nevertheless,
should all sta\-te\-ments of the accepted gravitomechanics be formally
 adopted for cone-masses,
the above derived field equations would
meet the problem of
 inhomogeneously curved 3D subspaces. It would not make
sense to speak about a common three-space for any ensemble
of intersecting extended objects.
It is impossible to introduce a common pseudo-Riemannian four-space
${\tilde x}$
with $g_{_1\mu\nu}({\tilde x})p_{_1}^\mu({\tilde x})
p_{_1}^{\nu}({\tilde x}) =
m_{_1}^2({\tilde x})$
and $g_{_2\mu\nu}({\tilde x})p_{_2}^\mu(\tilde x)p_{_2}^\nu(\tilde x) =
 m^2_{_2}({\tilde x}) $ for different
extended elementary mass densities at the same points of the same
manifold ${\tilde x}$,
because $g_{_1\mu\nu} ({\tilde x})\neq g_{_2\mu\nu}({\tilde x})$ due
 to different external systems for $m_{_1}$ and $m_{_2}$ at the same
  points. But extended intersections
of different cone objects may take place on  joint subspaces
when and if the latter hold common geometry.

Einstein's
covariant formalism can fluently operate, as known,  with different
 solutions for three-space metric under the pseudo-Riemannian four-interval.
Evolution of extended cone-particles can be observed through the dynamics
 of their point sources in
common material 3D space which should keep one universal geometry for all
 proper 3D subspaces. Observed conservation of three-momentum
 for mechanical systems at all space
points indicates that the common 3D space is to be homogeneous with
constant curvature (positive, negative or zero).

In order to verify mathematical  opportunities to implement
flat three-space into Einstein's scheme with pseudo-Riemannian
metric, we employ the known tetrad formalism,
 for example [8,23], which leads to the  representation of a four-interval,
  $ds^2 = g_{\mu \nu }dx^\mu dx^\nu $
 $\equiv \eta_{\alpha \beta  }e^\alpha_{\ \mu} e^\beta_{\ \nu}
  dx^\mu dx^\nu $ $\equiv \eta_{\alpha \beta}dx^\alpha dx^\beta $,
in the "plane" coordinates
 $dx^\alpha
\equiv e_{\ \mu} ^\alpha dx^\mu $, $dx^\beta
\equiv e_{\ \nu} ^\beta dx^\nu $,
 $\eta_{\alpha \beta }$
= $diag (+1,-1,-1,-1)$. One can immediately find
$e^o_{\ \mu} = \{  {\sqrt {g_{oo}}} ; - {\sqrt {g_{oo}}}g_i  \}$
and $e^a _{\ \mu} = \{0, e_{\ i}^a   \} $ from the equality
$ds^2 \equiv [{\sqrt {g_{oo}}} (dx^o-g_idx^i)]^2 - \gamma _{ij}dx^idx^j$.
At first glance the space triad $e_{\ i}^a$ (a = 1,2,3),
which  can be algebraically  represented via the components of the
three-space metric tensor $\gamma_{ij}$ $\equiv g_{oi}g_{oj}g^{-1}_{oo}
 - g_{ij}$, depends essentially on gravitation fields
or four-space distribution of  gravitomechanical cone-charges.
But this is not the case due to the universal degeneration of the
three-space
metric tensor $\gamma_{ij}$ for any elementary object.

Let us consider the   "curved" three-space components $V_i\equiv
{\cal V}_i
+ B^{\neq _N}_i$ of the metric-velocity four-vector $V_\mu \equiv
 g_{\mu \nu } {\dot x}^\nu $
 in (1) by
using the  tetrad formalism, $-({\sqrt {g_{oo}  }}g_i + v_i  )(1
- v_iv^i)^{-1/2} $
 = $V_i = e^\alpha _{\ i} V_\alpha = e^o_{\ i} V_o + e_{\ i}^a V_a $
= $-({\sqrt {g_{oo}  }}g_i + e_{\ i}^av_a  )(1 - v_av^a)^{-1/2} $.
This leads immediately to a trivial solution  $v_i \equiv e_{\ i}^av_a
 = \delta _i^av_a$
 for the "curved",  $v_i$,
and the "plane", $v_a$, three-velocities, because $
e^o_{\ i} =  - {\sqrt {g_{oo}}}g_i $ and $V_\alpha = \{
(1-v_av^a)^{-1/2}; - v_a(1-v_av^a)^{-1/2} \}$.
The solution with the   Kronecker delta-symbol $\delta _i^a$
($\delta^a_i=1, i=a$ and $\delta^a_i=0, i\neq a$)
demonstrates that the space triad
 and, consequently, the three-space metric
tensor  are  independent of gravitation
 fields, {\it i.e.} $e_{\ i}^a = \delta _i^a$ and
$\gamma _{ij}$ = $\delta _{ij}$.

Euclidean three-space geometry
may be appropriate for the covariant
formalism of gravitation due to the hidden equalities
$g_{oi}g_{oj}g^{-1}_{oo}$  $\equiv$  $g_{ij} + \delta_{ij}$
for every elementary material object.
 The metric tensor
in the most general case reads
$ g_{\mu \nu } \equiv
\eta_{\alpha \beta  } e^\alpha _{\ \mu} e^\beta _{\ \nu} \equiv
 \eta_{\mu \nu } +
 \eta_{\alpha \beta  } (e^\alpha _{\ \mu} e^\beta _{\ \nu} -
 \delta _\mu ^\alpha
 \delta _\nu ^\beta )\equiv \eta_{\mu\nu} + g_{\mu\nu}^{\neq _N}   ,$
where  $e^o_{\ \mu} = \{  {\sqrt {g_{oo}}} ; - {\sqrt {g_{oo}}}g_i  \}$
and $e^a _{\ \mu} = \{0, \delta _i^a   \} $ $\equiv \delta_\mu^a $.
In agreement
with this consideration,  the three-interval is
always associated with the universal Euclidean metric, because
$\gamma^{_N}_{ij} \equiv \{g_{oi}g_{oj}g_{oo}^{-1} - g_{ij} \}^{_N}
\equiv  \delta _{ij} \equiv - \eta_{ij}$ for all objects, while the
four-interval
is always associated with the proper pseudo-Riemannian metric,
$g_{\mu\nu}\neq \eta_{\mu\nu} $, which is different for different
 elementary objects.

A scalar differential of the four-interval along material
 points $x\equiv x_{_N}$ of any selected
cone object N (four-interval  $ds_{_N} \equiv ds$, for brevity) is  given by
\begin {equation}
ds^2 = \left [ {\sqrt {g_{oo}}} { dx^o\over ds }  + {{g_{oi}dx^i }
\over  {\sqrt {g_{oo}}} ds}    \right ]^2 ds^2
 - \delta_{i j} dx^idx^j \equiv d\tau^2 (ds,dt,dl) - dl^2
\end {equation}
 in arbitrary  external gravitational fields.
But (36) is a nonlinear equation with respect to $ds^2$, rather than
 a linear relation.
 The first term
at the right  hand side of (36) depends
on the four-interval   $ds$,
which is a nonlinear function of
 the 3D interval $dl \equiv dl_{_N} \equiv {\sqrt { \delta _{ij}dx^idx^j  }}
   $ and 1D interval $dt \equiv dt_{_N} \equiv {\sqrt { \delta _{oo}dx^odx^o
     }}  $.

Now we return to  the metric-velocity four-vector in (1).  Notice  that
$V_\mu = e^\alpha_{\ \mu} V_\alpha  = (e^a_{\ \mu} V_a + e^o_{\ \mu} V_o)$
= $ (e_{\ \mu}^a V_a  + \delta^o_{\mu} V_o ) $
+ $ (e_{\ \mu}^o - \delta^o_{\mu} ) V_o  $
$\equiv$ ${\cal V}_\mu + B^{\neq _N}_\mu$,  with the proper four-velocity
$ {\cal V}_\mu  $ $\equiv$
$(e_{\ \mu}^a V_a  + \delta^o_\mu V_o) = \delta_\mu^\alpha V_\alpha$,
 because $e^a_{\ o} = 0$ and $e^a_{\ i} = \delta^a_i$.
Flat three-space geometry
is just the way  to introduce the
 four-potential
$B^{\neq _N}_\mu  \equiv (e_{\ \mu} ^o   - \delta^o_\mu  )V_o $
 of external gravitational  field
in  analogy  with the four-potential $A_\mu ^{\neq _N}$
for external electromagnetic field.
This external  gravitational four-potential and the proper metric tensor
 $g_{\mu\nu}\equiv g^{_N}_{\mu\nu}$
are   characteristics of only one
 selected object N. In general the four-momentum density,
$P_{_N\mu }(x)^{s=o}_{x\neq \xi}
\equiv $ $ m_{_N} \delta^\alpha _\mu V_\alpha
+ m_{_N} B_\mu ^{\neq _N} + q_{_N} A^{\neq _N}_\mu$,
takes the mechanical, gravitational and
electromagnetic contributions, respectively.
Both gravitational and electromagnetic contributions
are associated with the same system of external
forming-up fields $a_{_K\mu}$, that provides for a neutral
object, $q_{_N}=0$, the following relations,
 \begin{eqnarray}
p_{_N\mu }(x) \equiv \left \{\!{{m_{_N}
  }\over {\sqrt {1 - v^2}}}+{{m_{_N} ({\sqrt {g_{oo}}}-1)
  }\over {\sqrt {1-v^2}}};
 -{{m_{_N} v_i}\over {\sqrt {1 - v^2}}}-{{m_{_N}g_i{\sqrt{g_{oo}}}
  }\over {\sqrt {1 - v^2}}}\right \}
\nonumber \\
\equiv
  m_{_N} \delta ^\alpha _\mu V_\alpha
+\sum_{_K}^{_K\neq _N}
(-G m_{_N}m_{_K}) a_{_K\mu }
 (x)^{s_{_K}=o}_{x\neq x_{_K}}.\end{eqnarray}

At the right hand side we used the symmetrical involvement
of any  mass, $m_{_K}$, and electric charge, $q_{_K}$, in their proper
gravitational and
electromagnetic field, based on the joint forming-up uncharged field
$a_{_K\mu}$.
This principle  statement of the developed scheme makes
external gravitational field linear with respect to the sources
and provides new opportunities to introduce a detail structure
of the  metric
tensor.
Both  gravitational, $B^{\neq _N}_\mu(x)\equiv - G \sum _{_K}^{_K\neq
 {_N}}\!m_{_K}
 a_{_K\mu}(x)^{s_{_K} =o}_{x\neq \xi_{_K}}$,
  and electromagnetic, $ A^{\neq _N}_\mu (x) \!\equiv$ $
\sum _{_K}^{_K\neq {_N}}\!q_{_K}
 a_{_K\mu} (x)^{s_{_K} =o}_{x\neq \xi_{_K}} $, four-potentials
may lead to the joint gauge-invariant scheme and to conservations of
the charges, $m_{_N}$ and $ q_{_N}$, res\-pec\-tively.
One will recall  that the mechanical  and the  gravitational charges in (37)
are equal.

In this section we study only neutral objects ($q_{_N} = 0$
for the selected object N) by staying in the framework of the mechanical
covariant formalism.
Then the tetrad  takes, according to (37),  the following components
$e_{\ \mu} ^a = \{0,   \delta_i^a \}$
$= \delta^a_\mu$ and $e^o_{\ \mu} =
\{1 +   {\sqrt {1-v^2} } B^{\neq _N}_o ;
 {\sqrt {1-v^2} }B^{\neq _N}_i  \} $
 $= \delta^o_\mu +  {\sqrt {1-v^2} }B^{\neq _N}_\mu    $,
and the proper  metric tensor $g^{_N}_{\mu \nu } \equiv \eta_{\mu\nu}
  + g_{\mu\nu}^{\neq _N}$
for the elementary cone-charge $m_{_N}$ in external fields  is given  by
\begin{equation}
\cases  { g_{oo} = (1 +   {\sqrt {1-v^2} }B^{\neq _N}_o )^2   \cr
       g_{oi} = (1 +   {\sqrt {1-v^2} }B^{\neq _N}_o )
 {\sqrt {1-v^2} }B^{\neq _N}_i    \cr
 g_{ij} = ({{1-v^2} })B^{\neq _N}_iB^{\neq _N}_j   + \eta_{ij}   \cr
}.
\end{equation}

 The  considered point $x$ of the selected object N is
 affected by all other objects K with
 retarded zero-interval signals from
their source point $\xi _{_K} (\tau_{_K})$.
As  expected, all components of the three-space metric
 tensor  are independent of  external gravitational charges,
$\gamma_{ij} \equiv g_{oi}g_{oj}g_{oo}^{-1} - g_{ij} = \delta _{ij}$
(now may be verified from (38)),
while every particular component of the proper four-space
metric tensor (38) is related to the external potential $B^{\neq
_N}_\mu(x)$.

One can represent (38) in a more compact way,
$g_{\mu\nu} = \eta_{\alpha\beta}e^\alpha_{\ \mu} e^\beta_{\ \nu}$
and    $ e_{\ \mu}^\alpha = \delta^\alpha_\mu + \delta^{\alpha o}
 {\sqrt {1-v^2} }B^{\neq _N}_\mu    $.
Notice, that
$P_{_N}^{\mu} = g^{\mu\nu}P_{_N\nu}$ = $m_{_N}[\eta^{\mu\nu}{\cal V}_\nu
 + (g^{\mu\nu} -
\eta^{\mu\nu}){\cal V}_\nu + g^{\mu\nu}B_\nu^{\neq _N}   ]$
= $m_{_N}\eta^{\mu\nu}{\cal V}_\nu$ - $m_{_N} \{ (1 + {\sqrt
 {1-v^2}}B^{\neq _N}_o   )^{-1}
(B^{\neq _N}_o + B^{\neq _N}_i v^i); 0 \}$ = $m_{_N}(1-v^2)^{-1/2} \{1 +
(g_{oo}^{-1/2} - 1 + g_i v^i  ); v^i \}  $ and $P_{_N\mu}P_{_N}^{\mu}$ =
$m^2_{_N} ({\cal V}_\mu {\cal V}^\mu  + 0)$ = $m^2_{_N}$ for arbitrary
gravitational four-potential $B_{_N\mu}^{\neq _N}$.

Substituting  the metric tensor (38) into  (36), we
obtain a general equation for the proper four-interval,   $ds = ds_{_N}$,
 of any selected cone object N,
\begin {equation}
 ds^2 + dl^2 =  (d{x}^o + {\sqrt{1-v^2}}
B^{\neq _N}_\mu{\dot x}^\mu ds)^2,
\end {equation}
where  ${\dot x}^\mu  \equiv dx^\mu / ds  $
and      $ dl^2  \equiv \delta _{ij}dx^idx^j  $.
Each elementary forming-up potential $a_{_K\mu }(x)^{s_{_K}=o}_{x\neq
 \xi_{_K}} $
within the gravitational  four-potential
 $B^{\neq _N}_\mu (x)  $  satisfies
the Maxwell-type equation
 with the proper metric tensor,
determined by all external fields for $m_{_K}$.
 This elementary potential takes the static Newton - Coulomb components
 $\{  r^{-1}_{_K}  ; 0 \}$ in a local rest frame of this  object K.
The radius $r_{_K}\equiv r_{_K} (x,\xi _{_K} [{\tau_{_K} }])^{s_{_K}=o}_{x
\neq \xi_{_K}}
\neq  0 $   is associated with  the "material"
parameter $\tau _{_K}$ determined by  a  zero
four-space interval, $s_{_K}(x,\xi _{_K}[\tau _{_K}]) =0$.

Now we derive a
planetary perihelion precession in order to
test the four-interval equations (39) or (36) with the new structure
of the metric tensor (38), which is consistent with flat three-space,
 $\gamma _{ij} = \delta _{ij}$.
A bound system of distant
  external sources with every $r_{_K} \approx r \equiv u^{-1}  $
may be considered as a united source
(the Sun, for example) with an
 effective mass
$M$. Relativistic relations for the function $\alpha_{_N}(ds^2,dl^2) \equiv
 {\sqrt{1-v_{_N}^2}}
B^{\neq _N}_\mu{\dot x}_{_N}^\mu $ will be derived in the next section.
One may use in (39) the Newton-Coulomb potential $B_o = - GMr^{-1} \equiv
 \mu u \ll 1$  for non-relativistic motion,
when a  considered object N  (a planet
with a small mass $m_{_N} \ll M$ and velocity $v^2 = dl^2/d\tau^2 \ll 1$)
 moves in Sun's
rest frame, where $B_{i} = 0$.
The equation (39) for the proper four-interval $ds$ reads
\begin{eqnarray}
 ds^2(dt,dl) + dl^2  = d\tau^2(dt,dl,ds) =
dt^2 \left(1 -\mu u{\sqrt {1 - {{dl^2}\over d\tau^2} }}   \right)^2
\nonumber \\
\approx dt^2 \left ( (1-\mu u)^2 + \mu u (1-\mu u){{dl^2}\over d\tau^2}
   \right )
= dt^2(1-\mu u)^2 + dl^2 \mu u (1-\mu u)^{-1},
\end{eqnarray}
where we used $(dx^o)^2 = dt^2$ and
 $dl^2   \ll    d\tau^2$.

The mass dependent coefficient at
the  three-interval, $dl^2 = dr^2 + r^2d\varphi^2 $
= $u^{-4}du^2 + u^{-2}d\varphi^2$, does not mean
departure from Euclidean  three-space geometry
in gravitational fields. This coefficient is associated with the direct
 involvement of space replacement $dl$ into the
proper time rate $d\tau (dt,dl,ds)$ in agreement with the metric tensor
(38).
 Our gravitational
time dilation and the  proper time rate,
$d\tau = {{(1 - GMu)} }dt$ for $v^2 \ll 1$, coincides with
the general relativity time rate, ${\sqrt {(1-2GMu)} }dt$,
 in  weak fields. But there are no Schwarzschild's divergencies
in the four-interval (39) for strong fields, as will be proved below.

The Killing vectors and integrals of motion,
$(1 - \mu u)^{2} {d t / ds} = E = const   $ and
 $r^2{d \varphi / ds} = L =const$ (with $\vartheta = \pi /2 = const$),
  are well known
under  the four-interval  (40) with stationary
coefficients in strong fields,
for example [24]. By taking into account
these conservation laws in  (40) one obtains
an equation for a rosette motion of planets
under the above  restrictions  on their velocities,
\begin{equation}
{(1-2\mu u) L^{-2}} + {(1-3\mu u)({u'^2 + u^2})} = {{E^2}L^{-2}},
\end{equation}
where $u'\equiv du/d\varphi$ and $\mu u \ll 1$. now one may
differentiate (41) with respect to
the polar angle
$\varphi$,
\begin{equation}
u'' + u - {\mu\over L^2 } = {9\over 2} \mu  u^2 +
3\mu u'' u + {3\over 2}\mu  u'^2,
\end{equation}
by keeping only the oldest nonlinear terms. This equation may be
 solved in two steps, when a linear
 solution, $u_o = \mu L^{-2}(1 + \epsilon cos \varphi)  $,
 is substituted  into the nonlinear terms
 on the right hand side of (42).

The most important
correction (which  is summed over century
rotations of the planets) is related to the "resonance"
(proportional to $ \epsilon cos \varphi$) nonlinear terms.
 We therefore ignore
in (42) all corrections, apart from $u^2 \sim 2\mu ^2L^{-4}
\epsilon cos \varphi$
and $u'' u \sim - \mu ^2L^{-4}\epsilon cos \varphi$. Then the  approximate
equation for the rosette motion, $u'' + u - \mu L^{-2} \approx
6\mu ^3 L^{-4} \epsilon cos \varphi  $, leads   to the
accepted perihelion precession, $\Delta \varphi = 6\pi \mu ^2L^{-2}
\equiv 6\pi \mu / a (1-\epsilon ^2) $, which was originally derived
from the Schwarzschild metric for the curved three-space, for example
[23-26].

It is important to emphasize that the
 measured result, $\Delta \varphi$,  for the planet
 perihelion precession  in weak Sun fields
was derived  from the
nonlinear four-interval (40) under  the flat three-space, rather than
from the linear four-interval under the curved three-space.
The four-interval equation (39) may be used only for a rest-mass object,
while a similar equation for photons, $dl^2/d\tau^2 = n^{-2}$, is associated
with their slowness $n^{-1}$ in gravitational fields. By taking into account
that $n^{-1} = {\sqrt {g_{oo}}}$ in static fields, one can explain the
measured gravitational light bending by keeping flat 3D space (Appendix 2).

 Thus, Euclidean 3D sub-space
provides the alternative way to explain main gravitational phenomena,
 to construct  self-contained relativity,
  and to overcome the known conceptual difficulties [27], associated with
Schwarzschild's solutions.
Covariant form of basic equations can hold universal  flat three-space,
which remains common for all  material objects, contrary to their proper
four-dimensional manifolds.
 One may note that flat three-space
is able to remove the conventional  objection (three-space curvature)
against the hypothesis [28] of electromagnetic origin of gravity.
There are many other arguments why flat three-space is exclusively
attractive for advanced theories of space-time [29].

Below we shall study the proper canonical pseudo-Riemannian four-space
 and  predict a new phenomenon
(electromagnetic time dilation and compression), which
have never been proposed from the other couplings of gravity and
electrodynamics, including [30-32]. This phenomenon is available
 for prompt laboratory tests  and may be interesting for applications.

 \bigskip \bigskip
\noindent
{\large{\bf 7.  Four-space with electromechanical connections}}

\bigskip
\noindent  { 7.1.  Electromagnetic time compression and dilation}

\bigskip
The observable motion of matter is three-dimensional in spite
 of the fact that various high dimensional manifolds can be employed for
 self-consistent description of any selected object.
Geometries
of the proper high dimensional manifolds  differ from the universal
(for all objects) geometry of the common 3D subspace.
 The proper metric tensor $g_{\mu\nu} \equiv
\eta_{\alpha\beta}e^\alpha_{\ \mu} e^\beta_{\ \nu} \neq \eta_{\mu\nu}$
 of pseudo-Riemannian four-space
may take only nonzero components, but must always hold
(in the developed constructions)
Euclidean geometry for 3D subspace, due to the hidden degeneration,
$g_{oi}g_{oj}g^{-1}_{oo}  - g_{ij} \equiv
\gamma_{ij}$  = $\delta_{ij}$, for real matter.
 This scheme provides a  simple opportunity
 for  implication of  electric charges into
metrics of the proper  pseudo-Riemannian four-spaces.
Contrary to the conventional formalism of
 general relativity, there are only four degrees of freedom from ten
  components $g_{\mu\nu}$ due to the above six conditions for 3D metrics.
One ought to expect that only four new relations, rather than ten
 Einstein-type equations $T_{_N}^{\mu\nu} = 0$, may remain independent
  for both gravitation and electrodynamics in flat three-space.

In order to describe objects with electric charge and rest mass
we return to (1) and (37), but with the symmetrical contribution
of gravitomechanical and electric charges into the  proper canonical
four-momentum,

\begin {eqnarray} P_{_N\mu }(x)
\equiv \left \{\!{{m_{_N}
  }\over {\sqrt {1 - v^2}}}+{{m_{_N} ({\sqrt {g_{oo}}}-1)
  }\over {\sqrt {1-v^2}}};
 -{{m_{_N} v_i}\over {\sqrt {1 - v^2}}}-{{m_{_N}g_i{\sqrt{g_{oo}}}
  }\over {\sqrt {1 - v^2}}}\right \}
\nonumber \\
\equiv m_{_N} V_\mu
 \equiv
  m_{_N} \delta ^\alpha _\mu V_\alpha
+\sum_{_K}^{_K\neq _N}
(-G m_{_N}m_{_K} + q_{_N}q_{_K} ) a_{_K\mu }
 (x)^{s_{_K}=o}_{x\neq x_{_K}}
\end {eqnarray}
or $P_{_N\mu }(x)^{s=o}_{x\neq \xi}$ $=$
$   \{ m_{_N} \delta^\alpha_\mu V_\alpha
+ m_{_N}U^{\neq _N}_\mu(x)\}^{s=o}_{x\neq \xi}$, with
$ \delta ^\alpha _\mu V_\alpha \equiv {\cal V}_\mu$ = $ \{
\beta^{-1}, - \beta^{-1}v_i \}$,   $U^{\neq _N}_\mu(x)
 $ = $  m^{-1}_{_N} \sum_{_K}^{_K\neq _N}
( -G m_{_N}m_{_K} + q_{_N}q_{_K} )a_{_K\mu }
 (x)^{s_{_K}=o}_{x\neq x_{_K}}$ =
 $B^{\neq _N}_\mu + m_{_N}^{-1}q_{_N}A^{\neq _N}_\mu$,
and $\beta$ = ${\sqrt {1 - \delta_{ij}v^iv^j}}$ = $ ds/d\tau
= ds / (ds^2 + dl^2)^{1/2} $.

By considering joint roots for the electric and gravitomechanical
 external fields, one can introduce (for a selected charged object N)
 a proper canonical four-dimensional space  ${x}_{_N}^\mu$ with
the affine connections generated by
 both electric and gravitomechanical external charges.
 The proper canonical four-momentum in this pseudo-Riemannian four-space
 takes the "old",
mechanical view,   $P_{_N}^\mu =
m_{_N}d{ x}_{_N}^\mu/ds_{_N}$, $ P_{_N\nu} =
m_{_N}{g}^{_N}_{\mu\nu}d{x}_{_N}^\mu/ds_{_N} =
m_{_N} \delta^\alpha_\nu V_\alpha + m_{_N}B_{\nu}^{\neq _N}
 + q_{_N}A_{\nu}^{\neq _N}$,
  $ds_{_N}^2$  = ${ g}^{_N}_{\mu\nu}d{ x}_{_N}^\mu
d{x}_{_N}^\nu $, $P_{_N\mu}P^{\mu}_{_N}$
  $ = m_{_N}^2$.
Then  all external electric charges
 contribute into the proper canonical metric tensor ${g}^{_N}_{\mu\nu} =
\eta_{\alpha\beta}e^\alpha_{_N\mu} e^\beta_{_N\nu}$ and into its tetrad
\begin {equation}
e^\alpha_{_N\mu}(x)^{s_{_N}=o}_{x\neq \xi_{_N}}  = \delta^\alpha_{\mu}
 + \delta^{\alpha o} {\sqrt {1-v^2}} U^{\neq {_N}}_\mu (x)^{s_{_N}=o}_{x
\neq \xi_{_N}}.
\end {equation}

One can verify from (44) that the canonical metric tensor is consistent
with the same flat three-space, ${g}_{oi}{ g}_{oj}
{g}^{-1}_{oo}  - { g}_{ij}$  = $\delta _{ij}$, as
well as the pure mechanical analog (37). The sole difference between the
proper four-intervals in the electromechanical and mechanical
 pseudo-Riemannian four-spaces is related to
the different proper times for charged and neutral objects in external
fields. But the proper times are always different even among electrically
 neutral
 objects and the additional contribution of
electrical charges into the proper time notion cannot change Einstein's
covariant formalism for relativistic motion.

In the general case the proper time of the selected charged object N,
$d\tau_{_N}\equiv d\tau = \beta^{-1} ds$ = ${\sqrt {g_{oo}}}(dx^o -
g_idx^i)$
 = $(1 + \beta U_o)d{x}^o
+ \beta U_id{x}^i = d{x}^o + \beta U_\mu d{x}^\mu   = d{x}^o + \beta^2 U_\mu
P^\mu m^{-1} d\tau $, depends on  all external gravitational and
 electromagnetic fields,

\begin {equation}
{{d \tau}}^2  = \left ( {  {  1 + \beta U^{\neq _N}_o  }\over
 {1 - \beta U^{\neq _N}_i v_{_N}^i } }\right )^2 dt^2
= {{dt^2}\over (1 - \beta^2 U_\mu^{\neq {_N}}P^\mu m^{-1})^2 }.
\end {equation}

Again, any curved proper four-space (which is specific for every neutral
 or charged object) is only auxiliary notion. Evolution of matter takes
  place
in three-space with the universal Euclidean geometry for all
extended charges and masses. There is no common curved Universe or
curved four-space for different 4D cone objects. Nevertheless
 one  may consider a common space+time manifold $\{dt({\bf x}); {\bf x}\}$
due to  the universal one-dimensional interval (7),
$dt^2_{_K} = { {\gamma^{_K}_{oo} dx^o_{_K}dx^o_{_K} }} $ with
$\gamma_{oo}^{_K} = \delta_{oo} = 1 $, and due to the common 3D
subspace, which keeps three-interval $dl^2_{_K} = \gamma^{_K}_{ij}
dx^i_{_K}dx^j_{_K}$ with the universal tensor $\gamma_{ij}^{_K} =
\delta_{ij}$   for all charged objects.

The physical velocities of charged cone objects in the flat three-space,
 $d{x}^i/d\tau$ $ \equiv v^i_{_N} \equiv v^i \equiv v_i \equiv v_{_Ni} $,
   are related with the proper time and, consequently,  with
 a particular distribution of external
sources of charges and masses. Note, that masses contribute into
 the external canonical potential $U^{\neq _N}_{\mu}$ with the same
sign and they can provide only the Einstein-type time dilation [1] in (45).
External electric charges with  different signs can lead to the
electromagnetic time compression, as well as to the electromagnetic
 time dilation. Both these phenomena are much
stronger for macroscopic charged objects than the gravitational time
dilation
 for these objects, and one may expect to test
the proposed electromagnetic compression-dilation of time
(45) in the laboratory.

There are experiments showing that the rate
of radioactive $\alpha$-decay may be accelerated by external electrical
 potential of the Van de Graaff generator [33].
There is also an observation [34] that tritium decays rapidly within the
 metal matrix that disagrees with the established theory of $\beta$-decay.
  The measured
decelerated oscillations of an electrically charged torque pendulum in a
 Faraday cage [35]
stimulate also to test (45) in practice.

By taking into account (45) with  $\beta = ds/ (ds^2 + dl^2)^{1/2}$, one may
relate all three proper intervals  $ds^2_{_N}\equiv {{
g^{_N}_{\mu\nu}dx_{_N}^\mu dx_{_N}^\nu}}$, $dl^2_{_N}
\equiv {{\delta_{ij}dx_{_N}^idx_{_N}^j}}$, and $dt^2_{_N}
 \equiv {{\delta_{oo} dx_{_N}^odx_{_N}^o}}$,
\begin {equation}
ds^2 + dl^2 = \left ( {{ds^2 + dl^2 + ds {\sqrt {ds^2 + dl^2}} U_o   }\over
ds^2 + dl^2 - U_i dx^i ds}   \right )dt^2.
\end {equation}

 The  relation (46) for the four-interval $ds \equiv ds_{_N}$
may be represented, due to the equalities
$ds^2_{_N} + dl^2_{_N} \equiv (dx_{_N}^o
+ \alpha_{_N} ds_{_N} )^2$ and $\alpha_{_N}  \equiv  \beta
 U_{\mu}^{\neq _N}P_{_N\mu}m^{-1}_{_N}$
$\equiv (U_o^{\neq _N} + U^{\neq _N}_i v^i )/(1 + \beta U_o^{\neq _N})$,
in the following form,
\begin {equation}
ds^2  \equiv
  \left ( {{\alpha_{_N} dx^o \pm {\sqrt {(dx^o)^2 -
  dl^2(1-\alpha_{_N}^2)} } }
\over (1 - \alpha_{_N}^2)}   \right )^2
 \approx {{dt^2}\over (1 - \alpha_{_N})^2}
- {{dl^2}\over (1-\alpha_{_N})},
\end {equation}
when $(1-\alpha_{_N}^2)dl^2/dt^2 \ll 1 $.
Notice that $\alpha_{_N} = \alpha_{_N}(ds, dt, dl)$
and there is no
Schwarzschild's peculiarity in the nonlinear equations (46)-(47)
 that provides a novel approach to the black hole problem.

The following summary of the main relations between the  proper
metric tensor, $g_{\mu\nu} \equiv g^{_N}_{\mu\nu}(x_{_N})$, the
 proper canonical
four-momentum, $P_{\mu}\equiv P_{_N\mu}(x_{_N})$,  and external fields,
$U_\mu \equiv U^{\neq _N}_\mu (x_{_N})$,    may be useful
for applications,
\begin{eqnarray}
  {g}_{oo} = (1 +  \beta U_o )^2,
    \   {g}_{oi} = (1 +   \beta U_o )
 \beta U_i, \
 { g}_{ij} = {\beta^2 }U_iU_j   - \delta_{ij},
\nonumber \\
{ g}^i = - { g}^{oi} = \gamma^{ij} { g}_j = { g}_i = - { g}_{oi}
{ g}_{oo}^{-1} =
  -U_i(\beta^{-1} + U_o )^{-1}
\nonumber \\
 g^{oo} = g^{-1}_{oo} -
 g_i g^i
= (1 - \beta^2 U_iU_j \delta^{ij})(1 + \beta U_o)^{-2}, \
\gamma_{ij} = \gamma^{ij}=  -g^{ij} = \delta_{ij}
\nonumber \\
P_{\mu} = mg_{\mu\nu}{{dx^\nu}/ ds}
= m ( \delta^\alpha_\mu V_\alpha + U_{\mu} )
= mu_\mu + qA_{\mu} = mV_\mu
\nonumber \\
P_\mu = m \{\beta^{-1} + U_o \ ; \ -\beta^{-1}v_i + U_i \}
   = m(\delta_\mu^\alpha V_\alpha + U_\mu) =
g_{\mu\nu}P^\nu
\nonumber \\
P^\mu =\{ m (\beta^{-1} + U^o) \ ; \ P^i  \} =  m \{\beta^{-1} -
(U_o + U_i v^i )(1 + \beta U_o)^{-1}\  ;\ \beta^{-1} v^i \}
\nonumber \\
 \ P_\mu P^\mu = g_{oo}
(P^o - g_i P^i)^2 - \delta_{ij}
 P^iP^j = P_o^2g^{-1}_{oo}
- m^2\beta^{-2} v^2    =  m^2.
\end{eqnarray}

It is remarkable that the contravariant component
$P_{_N}^i$ = $m_{_N}\beta^{-1} v^i$ does
 not depend on external potentials at all, $U_{\neq _N}^i \equiv 0$,
  contrary to
 the canonical three-momentum $P_{_N i }= - m_{_N}\beta^{-1}
 v_i + m_{_N}U^{\neq _N}_i$.
This means that the external electromagnetic potentials $A^{\neq _N}_\mu $
 and $A^\mu_{\neq _N}$, as well as the external gravitoelectromagnetic
  potentials $U^{\neq _N}_\mu, U_{\neq _N}^\mu$,
are not four-vectors in the proper four-space $x^\mu_{_N}$. The proper
 four-space
$x^\mu_{_N}$ and its proper metric tensor
$g^{_N}_{\mu\nu}$  were introduced for the proper notions of one selected
 object N, but not for its external fields. In our approach the proper
metric tensor
  cannot be applied for rising or lowering indexes of external
fields, {\it i.e.} $A^{\neq _N}_\mu \neq g^{_N}_{\mu\nu} A_{\neq _N}^\nu $
  and $U^{\neq _N}_\mu \neq g^{_N}_{\mu\nu} U_{\neq _N}^\nu $ (as well as
  $P_{_K\mu} \neq g^{_N}_{\mu\nu} P_{_K}^\nu $).
As a result, a pure electrical object with $q_{_N} \neq 0$ and
$m_{_N\mu} = 0$ can not exist in reality.

Note that
the conventional scalar product of any canonical four-vectors in classical
electrodynamics, for example
  ${g}^m_{\mu\nu}(mu^\mu + qA^\mu)
(mu^\nu + qA^\nu)$ = $ m^2$ + $2mq {g}^m_{\mu\nu} u^\mu A^\nu$
+ $q^2 {g}^m_{\mu\nu}A^\mu A^\nu $, is  not associated
with  conservations (because the pure mechanical metric
relations, ${g}^m_{\mu\nu}u^\mu u^\nu $ =1, are incorrect for  electrically
 charged objects). This fact  prevented to a reasonable introduction of
the  canonical four-momentum as an independent variable
in the classical theory,
which operates with the point particle in the collective field, rather than
with external fields  for the proper field densities
of every extended object.

Again, the proper metric tensors $g_{_N}^{\mu\nu}$ can not be applied
for rising indexes of any one summand in  $m_{_N}V_\mu = m_{_N}u_\mu +
 q_{_N} A_\mu^{\neq _N}$ = $m_{_N}{\cal V}_\mu + m_{_N}B_\mu^{\neq _N}
  + q_{_N}A_\mu^{\neq _N}$, despite $g^{\mu\nu}_{_N}V_\mu = V^\nu $ for
   the proper four-velocity  $V_\mu \equiv V_{_N\mu}$ of the charged
   object N.
Proper four-space and its geometry or metric tensor is associated only
with proper notions of a considered material object, rather with its
external fields $U^{\neq _N}_\mu = \sum^{_K\neq _N}_{_K} C_{_K} a_{_K\mu}$.
 There are proper tensors $g_{\mu\nu}^{_K}$, which relate the four-vectors
 $a_{_K\mu}$ and $a^\mu_{_K}$ in every four-space $x^\mu_{_K}$, but their
 is no common or universal metrics, which relates the quantities without
 definite tensor nature, $U_\mu^{\neq _N}$ and $U^\mu_{\neq _N}$.

The four-vector
 $P_{_N\mu} =  m_{_N}g_{\mu\nu}^{_N}{{dx^\nu_{_N}}/ ds_{_N}}
= m_{_N}u_\mu + q_{_N}A_{\mu}^{\neq _N}$
takes the unified structure, which cannot be informally split in
the proper four-space $x^\mu_{_N}$ into independent mechanical and
 electromagnetic four-vectors. Neither of the two summands,
$m_{_N}u_\mu$ and $ q_{_N}A_{\mu}^{\neq _N}$ is a four-vector
in the canonical four-space. One may verify from (48) that
$m_{_N}u_\mu m_{_N}u^\mu \neq m^2_{_N}$ for any electrically charged
particle in external electromagnetic fields. It is important
to underline, that the action (3b), for example, is associated with
the scalar product of the two four-vectors in the proper canonical
four-space $x^\mu_{_N}$, because $P_{_N\mu}i_{_N}^\mu \rightarrow
(m_{_N}g^{_N}_{\mu\nu}dx^\nu_{_N}/ds_{_N}) dx_{_N}^\mu $ $= m_{_N}ds_{_N}$.

Contrary to different geometries of curved four-spaces, Euclidean geometry
for flat 3D sub-spaces may be universally introduced for all
objects. This opens a way for comparisons, measurements and observations
 of different matter evolution in common 3D space. In actual practice,
 one may measure
 only a flat 3D interval, $dl = {\sqrt { \gamma^{_N}_{ij} dx^i d x^j  }} $,
and a flat 1D interval, $dt = {\sqrt { \gamma^{_N}_{oo} dx^o d x^o  }} $,
due to the universal relations  $\gamma^{_N}_{ij} = \delta_{ij}$
and $\gamma^{_N}_{oo} = \delta_{oo}$ for all objects. One cannot measure
a 4D interval $ds_{_N} = {\sqrt { g^{_N}_{\mu\nu} dx^\mu d x^\nu  }} $, as
well as $u_\mu$ or $V_\mu$, because $g^{_N}_{\mu\nu}$ for measured object
N does not coincides, $g^{_N}_{\mu\nu}\neq g^{_K}_{\mu\nu}$,
with metrics for other objects or instruments. In other words a
curved four-space space $x^\mu_{_N}$ is an auxiliary notion, and
 one may measure only $dl/dt$, rather than the mechanical four-velocity
  $u_{\mu}$
or canonical four-velocity $V_\mu$.

Now a natural question arises: Does equivalence principle take place in the
proper four-space with unified electromechanical connections?

\bigskip
\noindent  { 7.2.  Equivalence principle for charges}

\bigskip
Virtual and material variations $\delta x^\mu$ in the action (3a)
 lead to geodesic conservations of the canonical four-momentum $P_{_N\mu}$
  at all material points of the selected object N,

\begin {equation}
DP_{_N\mu}(x)^{s[\tau]=o}_{x\neq\xi[\tau]} = 0.
\end {equation}

These geodesic equations correspond to the equivalence principle
for the charged object N in its proper four-space with electromechanical
 connections. One may rewrite (49) in an equivalent form,
\begin {equation}
\{ P_{_N}^\nu(x) W_{_N\nu\mu} (x) \}^{s[\tau]=o}_{x\neq\xi[\tau]}
 = 0,
\end {equation}
because $P^\nu_{_N}\nabla_\mu P_{_N\nu} = 0$ and ${{D P_{_N\mu} }}
 \equiv {{dx_{_N}^\nu \nabla_\nu P_{_N\mu}  }}
= {{ds_{_N}m_{_N}^{-1}P_{_N}^\nu \nabla_\nu P_{_N\mu}  }}
= {{ds_{_N}m_{_N}^{-1}P_{_N}^\nu W^{_N}_{\nu\mu}  }}
 = 0$  at all
material points, $s[\tau] = 0$ and $x \neq \xi_{_N}$.
Now one may use the formal separation of the canonical tensor (14)
 into the three non-tensor
summands in these geodesic
equations, $P^j(\partial_j P_o - \partial_o P_j) = 0$ and
 $P^j(\partial_j P_i - \partial_i P_j) +
P^o(\partial_o P_i - \partial_i P_o) = 0$, in order to separate
the gravitational and electromagnetic forces under free motion of charges,
$$ P^\mu_{_N}(\partial_\nu {\cal V}_\mu - \partial_\mu
{\cal V}_\nu)  =  P^\mu_{_N}(\partial_\mu B^{\neq _N}_{\nu}
- \partial_\nu B^{\neq _N}_{\mu})  +
q_{_N}m_{_N}^{-1}P^\mu_{_N}(\partial_\mu A^{\neq _N}_{\nu}
- \partial_\nu A^{\neq _N}_{\mu}), \eqno (50')
$$
with ${\cal V}_\mu = \delta_\mu^\alpha V_\alpha = \{ \beta^{-1},
-\beta^{-1}v_i \} $. This is a replacement of the Minkowski-Lorentz
equation,
 $-m_{_N}Du^\mu / ds$ $\equiv$ $ m_{_N}u^\mu(\nabla_\nu u_\mu
- \nabla_\mu u_\nu) =  q_{_N}u^\mu F^{\neq _N}_{\mu\nu} $, for
electric charges in the conventional four-space with pure gravitomechanical
 connections, when $u_\mu u^\mu=1$.
 The Lorentz force, $q_{_N}m_{_N}^{-1}P^\mu_{_N}F_{\mu\nu}^{\neq _N}(x)$,
is accompanied  by its gravitational analog for masses,
$ P^\mu_{_N}G_{\mu\nu}^{\neq _N}(x)$, in (50'), where
$q(E + v\times B)_i + m(E_g + v\times B_g)_i$ =
$m\beta^{-1}[\partial_i {\cal V}_o - \partial_o {\cal V}_i - {\cal V}^j
(\partial_j {\cal V}_i - \partial_i {\cal V}_j)
 ]  $ $\approx m [\partial_o {v}_i + (v^j\nabla_j)v_i ] = m dv_i/dt$
and $m_{_N}v^i dv_i / dt = v^i (m_{_N}E_{gi} + q_{_N}E_i)$ for
 the nonrelativistic limit. There are no formal discrepancies
 with Newton's and
Lorentz's forces in this limit, but the self-action forces,
$P^\mu_{_N}m_{_N}f_{_N\mu\nu}(x)$ and   $q^2_{_N}m_{_N}^{-1}
P^\mu_{_N}f_{_N\mu\nu}(x)$, are absent under the geodesic motion
 (49)-(50'). The absence of self-action or self-acceleration  of
  free charges
in practice corresponds to restrictions $P^\mu_{_N}f_{_N\mu\nu}
= 0$ or  $dx^\mu_{_N}f_{_N\mu\nu} = 0$, which
can be employed in the present scheme along with the geodesic
relations (50).

\bigskip
\noindent {{ 7.3. Maxwell-type equation for vector electrogravity}}

\bigskip
The energy-tensor
 $T_{_N}^{\mu\nu}(x)_{x\neq \xi[\tau ]}^{s[\tau ]= 0}$ for the material
  cone-object is defined by (29) with $dx^\mu_{_N}/ds_{_N} = P^\mu_{_N}$,
$$
 T_{_N}^{\mu\nu}
\equiv
  \{ {{P_{_N}^{\mu}  I_{_N}^{\nu }
+ P_{_N}^{\nu}I_{_N}^{\mu } }\over 2}
 + {  W_{_N\rho\lambda}
\over 16 \pi}
[ g_{_N}^{\mu\nu} {f}_{_N}^{\rho\lambda} -
2g_{_N}^{\mu\rho}{f}_{_N}^{\nu\lambda}
- 2g_{_N}^{\nu\rho}{f}_{_N}^{\mu\lambda}] \}_{x\neq \xi[\tau ]}^{s[\tau ]
= 0}.
  \eqno (29')$$
Should this energy-tensor take only zero components as
it may be accepted for the conventional case, based on the supposed,
not proven, independency
of ten components in $g^{_N}_{\mu\nu}$?
Now one may exam  (from (48), for example) the relations between
components $P_{_N\mu}$ and $g^{_N}_{\mu\nu}$, with $P_{_N\mu}P_{_N\nu}
g^{\mu\nu}_{_N} = m_{_N}^2$. In the general case both  $P_{_N\mu}$ and
 $g^{_N}_{\mu\nu}$,
or the proper tetrad (44), depend on the same external
four-field $U_{\mu}^{\neq _N}$, while
  $ I_{_N}^\mu \delta P_{_N\mu } $ = $m_{_N}I_{_N}^\mu\delta
[ g_{\mu\nu}dx^\nu (g_{\rho\lambda}dx^\rho dx^\lambda)^{-1/2} ]$
 = $( P_{_N}^\mu I_{_N}^\nu + P_{_N}^\nu I_{_N}^\mu )
\delta g_{\mu \nu }/2$ for arbitrary proper four-vector $I_{_N}^{\mu}$.
  First, the metric tensor cannot be irrelevant under variations of the
   action (3a) with respect to $P_{_N\mu}$ that was mentioned under
   derivation of the Maxwell-type equations (5).
Second, not all components of $g^{_N}_{\mu\nu}$ are independent from one
another, and the Hilbert variation for the Einstein-type equation,
$T^{\mu\nu}_{_N}= 0$, is not a well defined procedure.

 The first general set (15) of Euler-Lagrange equations was derived
  after the variations of the action over changes of the proper
  field $a_{_N\mu}$.  Variations resulting from changes of the
  external field
 $U_{\mu}^{\neq _N}$, or any of its summand $a_{_K\mu}$,
may lead, in principle, to the other set of  four Euler-Lagrange
 equations. But it is important to remind that external fields are
  not proper four-vectors in the
curved four-space $x_{_N}^\mu$ and they cannot be employed as variables
for the scalar action (3a).

 An appropriate proper variable
is the four-vector $P_{_N\mu}$, with $\delta P_{_N\mu}$
 $= \delta U_{\mu}^{\neq _N}$ in the frame of reference with
  $v_i = 0$ and $\beta = 1$. Note that the proper tetrad $e^o_{\ \mu}
   = e_{o\mu}$, (44), is proportional to the
proper  four-momentum (43) in this frame, that may be used under
variation of (3b),
\begin {eqnarray}
\delta S =  -\int d^4x {\sqrt {-g}} T_{_N}^{\mu\nu}
\delta (e^\alpha_{_N \mu} e^{_N}_{\alpha\nu})
= - \int d^4x{\sqrt {-g}} 2T_{_N}^{\mu\nu} e^{_N}_{\alpha \nu}
\delta (\delta^\alpha_{\mu} + \delta^{\alpha o} U^{\neq _N}_{\mu}  )
\nonumber \\
= -\int d^4x{\sqrt {-g}} 2T_{_N}^{\mu\nu} e^{_N}_{o \nu}
\delta U^{\neq _N}_{\mu}
= -\int d^4x{\sqrt {-g}} 2m^{-2}_{_N}T_{_N}^{\mu\nu} P_{_N \nu}
 \delta P_{_N\mu} = 0.
\end {eqnarray}

The action ought to be constant with respect to
changes of external fields and corresponding variations of proper
 variables. Twelve, from sixteen, variational components in (51) are
  absent, because $\delta (e^b_{\ \mu}) $  $\equiv \delta (\delta_{\mu}^b) $
   = 0 when $b$ = 1,2,3.
The variations in (51) lead to a vector Euler-Lagrange equation,
\begin {eqnarray}
 {{ T^{\mu\nu}_{_N}(x) P_{_N\mu} }}(x)^{s=o}_{x\neq \xi} = 0,
\end {eqnarray}
rather than to a tensor Einstein-type equation $T^{\mu\nu}_{_N}
(x)^{s=o}_{x\neq \xi} = 0$.
Due to its tensor nature the equation (52) is valid under transformations
to frames of references with $v_{i} \neq 0$.
By substituting (29') in (52), one finds a four-vector equation,
\begin {equation}
 \left [
   I_{_N}^\nu + P_{_N}^\nu{{P_{_N\mu}I_{_N}^\mu}\over m_{_N}^2 }
- {{W_{_N\rho\lambda} }\over 8\pi m_{_N}^2}
\left ( 2P_{_N}^{\rho}
f_{_N}^{\nu\lambda} + 2g_{_N}^{\nu\rho}P_{_N\mu}f_{_N}^{\mu\lambda}
 - P_{_N}^{\nu} f^{\rho\lambda}_{_N} \right )\right ]^{s=o}_{x\neq \xi} = 0,
\end {equation}
and its scalar contraction,
\begin {equation}
P_{_N\nu}I^\nu_{_N} (x)^{s=o}_{x\neq \xi}
=
{{W_{_N\rho\lambda}(x)}\over 16\pi m_{_N}^2}
[  {{ 4 P_{_N}^{\rho}P_{_N\nu}f_{_N}^{\nu\lambda}(x) }}
 - m_{_N}^2 f_{_N}^{\rho\lambda}(x)]^{s=o}_{x\neq \xi}.
\end {equation}

Now one may use the geodesic relations, $P_{_N}^{\mu}W_{_N\mu\nu} = 0$,
and the self-action restrictions, $P_{_N}^{\mu}f_{_N\mu\nu} = 0$,  in order
to rewrite (54),
\begin {equation}
 P_{_N\nu}I^\nu_{_N} (x)^{s=o}_{x\neq \xi}
=
- {{1}\over 16\pi} {{W_{_N\rho\lambda}(x)} }
   f_{_N}^{\rho\lambda}(x)^{s=o}_{x\neq \xi},
\end {equation}
and to derive a vector Maxwell-type  equation,
\begin {equation}
\left \{ i_{_N}^\nu (x) - {1\over 4\pi} {\nabla_\mu f^{\mu\nu}_{_N}}
(x) +  {{  W_{_N\rho\lambda}(x)
   f_{_N}^{\rho\lambda}(x) }\over 16\pi m^2_{_N}}  P_{_N}^\nu (x)
\right \}_{x\neq \xi[\tau ]}^{s[\tau ]=o } = 0
\end {equation}
or  $ \{I^\nu_{_N} - m^{-2}_{_N}P_{_N\mu}I^\mu_{_N}
P^\nu_{_N}\}^{s=o}_{x\neq \xi} $ = 0,
which generalizes the equation (5) for
the potential state N with $W_{_N\rho\lambda}(x)^{s=o}_{x\neq \xi} = 0$.

Conservation of the mass or electrical charge four-current,
 $q_{_N}\nabla_{\mu}i_{_N}^\mu = 0 $, is verified by practice, while
one finds $\nabla_\nu i_{_N}^\nu \equiv \nabla_\nu I^\nu_{_N}
 = m^{-2}_{_N} \nabla_{\nu}( P_{_N\mu}I^\mu_{_N} P^\nu_{_N}) $ from (56).
  In order to prove that (56) is consistent with the charge conservation,
   we use infinitesimal
 homogeneous coordinate shifts, $x'^\mu = x^\mu + \xi^\mu$, in the action
  (3a). This procedure, for example [8], leads to the
following vector conservations for the elementary cone object,
\begin {equation}
\nabla_\mu T^\mu _{_N\nu} = \left \{ {{\nabla_\mu \left (
P_{_N}^\mu I_{_N\nu}  + P_{_N\nu}I^\mu_{_N}
\right )}\over 2} -
 {{ W_{_N\nu\lambda} \nabla_\rho f_{_N}^{\rho\lambda}
+ f_{_N\nu\lambda} \nabla_\rho W_{_N}^{\rho\lambda}}\over
8\pi}  \right \}^{s={o}}_{x\neq \xi} = 0,
\end {equation}
where we used $W_{_N}^{\rho\lambda} \nabla_\nu f_{_N\rho\lambda}
= 2W_{_N}^{\mu\lambda} \nabla_\mu f_{_N\nu \lambda}$,
$f_{_N}^{\rho\lambda} \nabla_\nu W_{_N\rho\lambda}
= 2f_{_N}^{\mu\lambda} \nabla_\mu W_{_N\nu \lambda}$
due to (12) and (16). Contraction of (57) with $P_{_N}^\nu$, under (50)
and (15), reveals
 one more scalar condition under the geodesic motion
of free
cone-charges, $ m^2_{_N}
\nabla_\mu I^\mu_{_N}(x)^{s=o}_{x \neq \xi} = - P_{_N}^\nu \nabla_\mu
(P^\mu_{_N}I_{_N\nu} )^{s=o}_{x \neq \xi}
\equiv  - \nabla_\mu (P_{_N}^\mu P^\nu_{_N}I_{_N\nu} )^{s=o}_{x \neq \xi} $.
 This condition
is compatible with the covariant divergence of the vector (56),
$ \nabla_\mu I^\mu_{_N}(x)^{s=o}_{x \neq \xi}
=  m^{-2}_{_N} \nabla_\mu (P_{_N}^\mu P^\nu_{_N}I_{_N\nu}
)^{s=o}_{x \neq \xi} $, only under conservation of the cone-particle
 four-flow,
when $\nabla_\mu I^\mu_{_N}(x)^{s=o}_{x \neq \xi} \equiv  \nabla_\mu
 i^\mu_{_N}(x)^{s=o}_{x \neq \xi} = 0$ and
$\nabla_\mu (P_{_N}^\mu P^\nu_{_N}I_{_N\nu} )^{s=o}_{x \neq \xi} = 0 $.

 Potential or superfluid states without radiation or energy exchange
  with external matter obey the particular
 restrictions, $I_{_N}^\mu = W_{_N\mu\nu} =  T^{\mu\nu}_{_N}= 0$,
 which satisfy the
general equations (52)-(56). Both gravitation and electrodynamics of
 superfluid charges and their potential states, based on (5), (12),
  (15), (16) and (30), may be represented in a gauge invariant form
   with respect to the proper four-vectors $a_{_N\mu}(x) \rightarrow
    a_{_N\mu}(x) + \partial_\mu \chi_{_N}(x) $ and
$P_{_N\mu}(x) \rightarrow P_{_N\mu}(x) + \partial_\mu \Upsilon_{_N}(x) $,
with $\partial_\mu \partial_\nu \chi_{_N}(x) = \partial_\nu \partial_\mu
\chi_{_N}(x) $ and $\partial_\mu \partial_\nu \Upsilon_{_N}(x)
 = \partial_\nu \partial_\mu \Upsilon_{_N}(x) $. Note that gauge
  transformations may be
introduced only for proper four-vectors, rather than for external fields
$B^{\neq _N}_\mu$  and  $A^{\neq _N}_\mu$, which are not four-vectors
in the considered four-space $x^\mu_{_N}$. Could one employ for a moment
 common metrics for all four-spaces $x_{_K}^\mu$ (pseudo-Euclidean common
four-space in classical electrodynamics, for example), gauge transformations
 might be considered for external potentials, as is known.

Radiation or energy exchange with external matter is not allowed for
the potential motion.
Non-potential motion
ought to satisfy the general equations (56), (12), (15), (16), and (50).
The energy-tensor density may have nonzero
components, but
 $T_{_N}^{\mu\nu}P_{_N\mu} = 0$ under general motion  in vector
  electrogravity. Emission and absorption of real photons, rather
  than gauge bosons, violate the gauge invariance, associated in
  the present scheme only with the proper fields under superfluid,
   potential states.

Different solutions of the general variational equations (15),
(49) and (56)  may be used to classify various forms of matter motion.
Both gravitational and electromagnetic waves are associated with vector
photons in the developed unification. These vector waves may modulate
the four-space metric tensor $g_{\mu\nu}$, but they are irrelevant to
Euclidean metrics, $\gamma_{ij} = \delta_{ij}$, of  three-space.
Vector electrogravity is consistent with the absence of
space modulations in all known experiments [26] and predicts
flat laboratory metrics for gravitational waves and all other relativistic
phenomena.

According to conventional theory the
electromagnetic and gravitational parts of interactions cannot compensate
each other in the state of general motion due to the different
tensor nature of these interactions. But interactions with external
 gravitational
and electric charges are determined by the same term  $U_\mu^{\neq _N}
 = m^{-1}_{_N}
\sum_{_K}^{_K\neq _N} (q_{_N}q_{_K} - G m_{_N}m_{_K})a_{_K\mu}$
in the developed vector unification. The masses $m_{_K}$ with their
vector fields $a_{_K\mu}$, rather than the energy tensor density,
are sources of gravity. The universal compensation of external
fields $B_\mu^{\neq _N}$ and $A_\mu^{\neq _N}$ for a two-body system
with $q_{_1}q_{_2} = Gm_{_1}m_{_2}$, for example, is drastically different
from the case of general relativity. The novel, vector field nature
of gravity
in VEG, is consistent with the attractive opportunity [28] of
 electromagnetic origin of the extended mass $m_{_K}$, that may be studied
under further developments of the present theory.

\bigskip \bigskip

\noindent {\large {\bf 8. Conclusion}}

\bigskip

Our non-dualistic   approach   to the  extended
object was initiated by the introduction of the elementary cone-particle
and the elementary cone-field
in terms of a multifractional field
 emanating from one point source.
Every object N with a rest-mass $m_{_N}$  contains
  its proper forming-up field $a_{_K\mu} (x)$ at all points  of the proper
 four-space $x^\mu = x_{_N}^\mu$,   which are related by
 zero four-intervals with respect to each other and the joint point
vertex $\xi $. Proper fields $a_{_K\mu} (x)^{s_{_N} =o}_{x\neq \xi_{_N}}$
of other extended objects K are external
for the selected object N when they cross its family of charged material
points, called as a light cone.

The proper four-space,
   $dx_{_N}^\mu = g_{_N}^{\mu\nu} dx^{_N}_\nu$, with electromechanical
connections ought to be personally introduced for every charged object.
 Pseudo-Riemannian metric tensor of this proper four-space,
 $g^{_N}_{\mu\nu}(x)  = \eta_{\alpha\beta}e^\alpha_{_N \mu}
 e^\beta_{_N \nu} $, with  $e^\alpha_{_N \mu} = \delta^\alpha_\mu
  + \delta^{\alpha o}{\sqrt {1-v_{_N}^2}}   \sum _{_K}^{_K\neq {_N}}
   (-Gm_{_K} + m^{-1}_{_N}q_{_N}q_{_K})
 a_{_K\mu} (x )^{s_{_K} =o}_{x\neq \xi_{_K}}  $,
$g_{oi}g_{oj}g^{-1}_{oo}  - g_{ij} \equiv
\gamma _{ij}$  = $\delta _{ij}$,  ${\sqrt {-g}}  = {\sqrt {g_{oo}}}$,
 is determined by local densities of charged fields generated by distant
  external sources. These densities of external fields contribute to the
   proper canonical four-momentum density of the selected object,
$P_{_N\mu}(x)^{s=o}_{x\neq \xi}$  $\equiv  m_{_N}g^{_N}_{\mu\nu}
dx^\mu/ds$  = $m_{_N}\delta^\alpha_\mu V_\alpha $ +
$m_{_N}\sum _{_K}^{_K\neq {_N}} (-Gm_{_K} + m^{-1}_{_N}q_{_N}q_{_K})
 a_{_K\mu} (x )^{s_{_K} =o}_{x\neq \xi_{_K}} $. The action of the
cone-object depends on the vorticity tensors  $W_{_N\mu \nu }(x)
= \partial_\mu P_{_N\nu}(x)
- \partial_\nu P_{_N\mu}(x)$ and
$f_{_N\mu \nu }(x) = \partial_\mu a_{_N\nu}(x)
- \partial_\nu a_{_N\mu}(x)$ at all proper
material points $x = x_{_N}$, with  $x\neq \xi [\tau]$ and
$s(x,\xi[\tau ])=0$.

Mirror cone particles K$_{_1}$ and K$_{_2}$ may be described
by
introducing
 the opposite parametric differentials $dt_{_1,_2}$
at every point of three-space {\bf x}. Both mirror cones for matter and
 for antimatter with one joint vertex
(excluded from  material cone states) in four-space $\{x^o;x^i\}$  contain
 their own
field and particle fractions.
The elementary cone-field and the elementary cone-particle are locally
bound at every material
 point of their joint geometrical structure.
One can apply two mirror
space+time manifolds, $\{dt_{_1}; {\bf x}\}$ and
$\{dt_{_2}; {\bf x}\}$
with  $dt_{_1}=-dt_{_2}$,
 for symmetrical evolution of matter and
antimatter (in agreement with CPT symmetry) under  only
 retarded their
emission from all point sources.

Two tensor equalities (12), (16)  and five vector equations
$\nabla_{\mu}W_{_N}^{\mu\nu}(x)^{s_{_N} =o}_{x\neq \xi_{_N}} $ =
$P_{_N\mu}T^{\mu\nu}_{_N}(x)^{s_{_N} =o}_{x\neq \xi_{_N}}$ =
$P_{_N}^{\mu}W_{_N\mu\nu}(x)^{s_{_N} =o}_{x\neq \xi_{_N}}$ =
$P_{_N}^{\mu}f_{_N\mu\nu}(x)^{s_{_N} =o}_{x\neq \xi_{_N}}$ =
$\nabla_\mu T^\mu_{_N\nu}(x)^{s_{_N} =o}_{x\neq \xi_{_N}}$ =0
 may propose the novel description of charged objects in vector
  gravitational and electromagnetic fields.
The total scheme of vector electrogravity
is self-consistent and
 demonstrates
the  unified foundation for  elementary gravitational and electromagnetic
proper fields - they are associated with the constant field densities,
$m_{_N}$
 and $q_{_N}$,
on the basis of the same forming-up cone-field
$a_{_N\mu} (x)^{s_{_N} =o}_{x\neq \xi_{_N}}  $.
Both kinds of extended cone-charges cannot curve  Euclidean three-space,
 but they are responsible for the curved four-space and for the
proper time dilation-compression. The electric charge cannot
exist without the mass and  electromagnetic part of
the canonical four-momentum cannot be considered independently
as a proper four-vector.

Contrary to curved pseudo-Riemannian four-spaces $ x^\mu_{_N}$
with different proper metric tensors
$g_{\mu\nu}^{_N}$, all their proper 3D subspaces
 ${x^i}_{_N}$
exhibit universal Euclidean geometry, $dl^2_{_K} =
\gamma^{_K}_{ij}dx_{_K}^idx_{_K}^j$ and $\gamma^{_K}_{ij} = \delta_{ij}$.
 The proper
 time interval  is also independent from  proper
 characteristics
of different objects, $dt^2_{_K} = \gamma^{_K}_{oo}dx_{_K}^odx_{_K}^o$
 and $\gamma^{_K}_{oo} = \delta_{oo} = 1$. For this reason
 the common space+time manifold
 $ \{dt, { x^i}\} $  (not a joint curved four-space $\{x^o; {x}^i\}$
with the common metric tensor for all particles, like in general
 relativity) may be introduced for the description of evolution
 of all objects in the flat
three-dimensional Universe.

It is important to underline that vector electrogravity essentially
 differs from
general relativity in the description of gravity. All interactions
are linear in VEG and superposition principle is satisfied with
respect to the sources. The origin of gravity is the extended
mass in the present scheme, but not
the stress-energy tensor. The four-component
vector equation $T_{_N}^{\mu\nu}P_{_N\nu} = 0$, rather than the
 ten "independent" components of the tensor equation $T_{_N}^{\mu\nu}
  = 0$, is responsible for both gravitation and electrodynamics in the
   most general case.

Being unified with electrodynamics, vector gravitation of extended masses
 becomes  a self-contained
 theory, which may be applied to practice
without references of  other gravitomechanical theories.
Nevertheless, the developed electrodynamic approach to gravitation
coincides completely with Newton's theory in the nonrelativistic limit.
 The available observations
for all known kinds of interactions and conservations  do not
contradict the employed Euclidean 3D geometry for
the united space-charge-mass  continuum.

The developed linear  synthesis of external electromagnetic and
 gravitational fields
(under the nonlinear proper four-interval),
and the integration of the cone-particle into the very  cone-field
structure satisfy   the predicted double unified criterion [11], as
well as  the
other known criteria [20,21] for the unified field theory.
The nonmaterial point sources ({\it i.e.} peculiarities of matter)
 are excluded from the material field equations in  agreement with
 Einstein's approach [36] to the continuum theory, and all  physical
magnitudes of vector electrogravity are free from divergencies.

The advanced, but not retarded, emission of charged fields from point
 sources and self-acceleration
of extended charges are absent in electrogravity, contrary to classical
 electrodynamics. Both gravitation and electrodynamics of  superfluid
 charges,
based on (5), (15), and (49), take the gauge-invariant form.
Electrogravity of extended masses in flat three-space is consistent with
the hypothesis about
 the electromagnetic  origin of gravitation [28],
while  Schwarzschild's three-space curvature around the point particle
cannot be satisfactorily agreed with the electromagnetic nature of mass.

 As to experiments, all gravitational observations correspond to the
 introduced concept of flat 3D space and flat 1D time intervals
for 4D cone objects. The  extended electric
cone-charge (and cone-mass)  is consistent with the celebrated
Aharonov-Bohm phenomenon [15]
 and   the  relativistic experiments with rotating
superconductors [14]. A practical opportunity to test the proposed
 scheme for electrogravity in the laboratory is to
verify the predicted  electromagnetic  compression-dilation of time
 for charged objects,  like in
the experiments [33-35].

The introduced concept of the extended cone-charge rejects the
classical  three-space with bulk, particle free regions.
Material intersections of curved charged four-spaces on flat 3D subspaces
 may clarify  the "action-at-a-distance" and energy transfer within
 the common space-charge-mass continuum.
Vector nature of electrogravity corresponds to the unified,
 photonic way for all kinds of electromagnetic and gravitational
  radiations, and predicts the absence of metric modulations of
  flat, laboratory space under the forthcoming search of gravitational
   waves
[37].

\bigskip\bigskip
\noindent {\large{\bf Appendix 1: applications of the
 {${\hat \delta}$-operator }}}

\bigskip

The  ${\hat \delta }^4_{_N}(x,\xi )^{s=o}_{x\neq \xi }$-operator
 formalism for the conjugation of  the point source at $\xi$
with the infinite  material continuum at $s(x,\xi) = 0$ and $x\neq \xi$
may be demonstrated
for  flat four-space
($g^{\nu\nu}= 1, g^{ij} = \delta^{ij}$, $g_{ii}=-1, g_{oo}=1  $,
$d\xi_{\mu } = \eta _{\mu \nu }d\xi^\nu $ and
$a _{_N\mu }(x,\xi) = \eta _{\mu \nu }a^\nu _{_N}(x,\xi)$,
$\eta_{_N}(x,\xi)_{x\neq \xi} $
 $ = $ $ \eta_{_N} (|x-\xi|)_{x\neq \xi} $,  $|x-\xi|$ =
 $[(x^o - \xi^o)^2 - ({\bf x} - {\mbox {\boldmath $\xi $}})^2 ]^{1/2}$),
where the equation (5) can be  simplified,
$ \partial_\mu \partial^\mu a_{_N}^\nu
(x,\xi)^{s=_0}_{x\neq \xi}$  =
$4\pi { i}^\nu _{_N}(x,\xi)^{s=_0}_{x\neq \xi}$,
due to the subsidiary condition
$\partial^\mu \partial^\nu a_{_N\mu} \equiv
\partial^\nu \partial^\mu a_{_N\mu}
$ $ = $ $\partial^\nu \int d\xi_\mu$ $
\partial^\mu \eta_{_N} (|x - \xi[p]|)  $ $=$ $- \partial^\nu
\{ \eta_{_N} (|x - \xi[p]|) |^{p =+\infty}_{p=-\infty} \} $
 $ = 0$.

The simplified equation (5) may be solved via
Green's function $ { G}(x,x')_{x\neq x'}$ = $\{{ \delta }
(x^o-{x'}^o\mp|{\bf x}-{\bf x}'| ) /
|{\bf x}-{\bf x}'|\}_{{x}\neq{x}'}$, associated with equation
$ \partial_\mu \partial^\mu { G}(x,x')_{x\neq x'}$ = $
{\hat \delta}^4(x,x')_{x\neq x'} $ $\equiv$ $\{{\hat \delta}^3
({\bf x},{\bf x}')$${ \delta }
(x^o-{x'}^o\mp|{\bf x}-{\bf x}'| )\}_{x\neq x'}$. The direct and the
inverted
 elementary fields are associated with
different contravariant
four-vectors,
$$
\ \ \ \ \ \ \ \ a_{_N}^\nu (x)^{s[\tau _{_1,_2}]={_0}}_{x\neq \xi
[\tau _{_1,_2}]} =\int
 d^4 x' {{i^\nu_{_N}(x')^{s'[\tau _{_1,_2}]=_0}_{x'\neq\xi[\tau _{_1,_2}] }
 } }  { G}(x,x')_{x\neq x'}
$$ $$ 
=\!\int\!dp\!\int\!d^4 x'
 {{    { {dx'}^\nu\over dp  }\!{\hat
 \delta}_{_N}^4(x',\xi[p])_{x'\neq \xi}
} } { G}(x,x')_{x\neq x'}
=\!\int\!dp\!{{{{  d \xi^\nu[p] }\over dp}
{G}(x,\xi[p])_{x\neq \xi }
}}
$$ $$
= \int {{dp}\over
 |{\bf x}-{\mbox {\boldmath $\xi $}}[p]|_{{ x}\neq{ \xi}[p]}}
{{d \xi^\nu[p] }\over dp}
 {{    { \delta}
 (p - \tau _{_1,_2})
} \over \left |
 {{\partial \xi^o[p]}\over \partial p } \pm
   {{ \partial |{\bf x}- {\mbox {\boldmath $\xi $}}[p] |  }\over
\partial p}
\right |_{{ x}\neq
 { \xi}[p] } } =
$$ $$ 
  {{ {{ d\xi^\nu[\tau _{_1,_2}]/d\tau _{_1,_2}
} }
   }\over
 \left |r[\tau _{_1,_2}]
{{ \partial \xi^{_0}[\tau _{_1,_2}]
}\over \partial \tau _{_1,_2} }
 \mp {\bf r}[\tau _{_1,_2}]
{{ \partial {\mbox {\boldmath $\xi $}}[\tau _{_1,_2}] }\over
 \partial \tau _{_1,_2} }
    \right |^{s[\tau _{_1,_2}]={_0}}_{{\bf x}\neq {\mbox {\boldmath $\xi $}}
[\tau _{_1,_2}] }     }$$ $$
=
 {{ d\xi^\nu [\tau _{_1,_2}] }
\over |r[\tau _{_1,_2}]|  {{d\tau _{_1,_2}}\over |d\tau _{_1,_2}|}
 \left | d\xi_o[\tau _{_1,_2}] \left ( 1
\mp {{\bf r}[\tau _{_1,_2}]\over r[\tau _{_1,_2}] }
{{ d {\mbox {\boldmath $\xi $}}[\tau _{_1,_2}] }\over
 d \xi^o [\tau _{_1,_2}] }
    \right ) \right |_{{\bf x}\neq
{\mbox {\boldmath $\xi $}}[\tau _{_1,_2}]}^{s
[\tau _{_1,_2}]={_0}} }$$
$$ =  {{ d\xi^\nu [\tau _{_1,_2}]  }
\over |r[\tau _{_1,_2}]|  {{d\tau _{_1,_2}}\over |d\tau _{_1,_2}|}
 \left | d\xi_o[\tau _{_1,_2}] \mp
d| {\mbox {\boldmath $\xi $}}[\tau _{_1,_2}] - {\bf x}|
     \right |_{{\bf x}\neq
{\mbox {\boldmath $\xi $}}[\tau _{_1,_2}]}^{s
[\tau _{_1,_2}]={_0}} }
=  {{d\xi^\nu [\tau _{_1,_2}]}\over dt_{_1,_2}
 |r[\tau _{_1,_2}]|_{{\bf x}\neq
{\mbox {\boldmath $\xi $}}[\tau _{_1,_2}]}^{s
[\tau _{_1,_2}]={_0}} }    , $$
where we used
${\bf r}[\tau ]d{\mbox {\boldmath $\xi $}}[\tau ] $
= $- { r}^i[\tau ]d{\xi}_{i}[\tau ] $,
$r^i[\tau] = x^i - \xi^i[\tau] $,
 $|{\bf x } - {\mbox {\boldmath $\xi $}}[\tau ] |$ $ \equiv$ $r[\tau]  $
 $= {\sqrt {- r_{i}[\tau ]r^i[\tau]}} $,
$\xi^{_0}[\tau _{_1}] = x^{_0} - r[\tau _{_1}]   $,
$\xi^{_0}[\tau _{_2}] = x^{_0} + r[\tau _{_2}]   $, and
${ \delta} [f(p)]$ = $ {\delta}
 (p-\tau )/ |f'(p)|$
with $f(\tau )= 0$.
Two different four-potentials
 $a_{_N}^\nu  (x)^{s[\tau _{_1,_2}]={_0}}_{x\neq \xi[\tau _{_1,_2}]}$
can be represented in
 the  Lienard-Wie\-chert
form with the velocity
$  d\xi^{\nu}[\tau _{_1,_2}] / d t_{_1,_2}  $
of  point sources of matter and antimatter, respectively,
in the mirror  space-time manifolds.

In order to derive the covariant  four-potentials
from their definition,  one should
solve (8), $ d\xi^{\nu}[p]\partial_\mu\partial^\mu
 { \eta}_{_N}(|x-\xi [p]|)_{x\neq \xi}
 = 4\pi dx^\nu {\hat \delta}_{_N}^4(x,\xi [p])_{x\neq \xi}
 $ (in  flat four-space), via the same Green's function
${G}(x,x')_{x\neq x'}$,
$$ { \eta}_{_N}(|x-\xi[p]|)_{x\neq \xi[p]}
 =
\int d^4  x'
 {\hat \delta}_{_N}^4(x', \xi[p])_{x'\neq \xi[p]}
 {G}(x,x')_{x\neq x'}
 {{dx'^\nu   }
\over d\xi^{\nu}[p] }$$
$$\ \ \ \ \ = { G}(x,\xi[p])_{x\neq \xi[p]} =
\left \{   {{ { \delta }
(x^o-\xi^o[p]\mp|{\bf x}-{\mbox {\boldmath $\xi $}}[p]|
) }
\over |{\bf x}-{\mbox {\boldmath $\xi $}}[p]| }
\right \}_{x\neq \xi[p]}.  $$

By using this relation directly in the definition of  the
  basic covariant four-potential,
$ a_{_N\nu}(x)^{s[\tau]={_0}}_{x\neq \xi[\tau]}$ = $
 \int  dp  { G}(x,\xi[p])_{x\neq \xi[p]}
 \{ {{d{\xi}_{\nu}[p]}}
/d p\}$,
 one can verify  the
  above derived result for both the direct
 (for matter) and the inverted (for antimatter)
solutions  with $dt_{_1}$  and $dt_{_2}$, respectively.

\bigskip\bigskip
\noindent {\large{\bf Appendix 2: Gravitational light bending}}

\bigskip

A four-interval equation for electromagnetic waves in nonstationary
gravitational fields may be derived
from the light velocity, $dl/d\tau  = n^{-1}$ or
$$
\delta_{ij} dx^i dx^j = n^{-2}g_{oo}(dx^o - g_idx^i)^2,
$$
where the slowness $n^{-1} \equiv 1 / n(g_{oo}, g_i)$ is associated with
the metric contribution
into the Maxwell and wave equations.
This  slowness defines  the wave and ray equations,
$k_\mu k^\mu = (1-n^2)(k_\mu V^\mu)^2$ and $\delta \int k_\mu dx^\mu = 0$,
for light in arbitrary gravitational fields [38].

Below we consider for simplicity only a static gravitational
field. It formally works like a static medium for electromagnetic waves,
 because  ${\bf D} = g_{oo}^{-1/2} {\bf E} $, ${\bf B} = g_{oo}^{-1/2}
 {\bf H} $, and   $n^{-1} = ({ {  { \tilde \varepsilon} {\tilde \mu}}}
 )^{-{1/2}} =
{{{{g^{1/2}_{oo}}} }}$
in the covariant Maxwell and wave equations [8].

Equations for light rays in static gravitational fields may be found
from Fermat's principle [38],
$$
\delta\!\int k_i dx^i = -\delta\!\int\!  {{k_o dl}\over  n^{-1}{\sqrt
 {g_{oo}}}} =
-k_o\delta\!\int {{dl}\over {{g_{oo} }} }
= -k_o\delta\!\int {  { \sqrt {
  du^2 + u^2d\varphi^2   }   }\over u^2 (1- GM u)^2   } = 0,
$$
where $n^2 k_o k^o = k_i k^i,  {{g_{oo}}}k^o
= k_o = const $,  $k_i = -nk_o g_{oo}^{-1/2}
dx^i/dl$,  ${\sqrt {g_{oo}}}= 1-GM u$,
$g_i=-g_{oi}/g_{oo}=0$,
$dl = {\sqrt { \delta_{ij} x^ix^j}}$  = $
{\sqrt {       dr^2 + r^2d\varphi^2}}$
($ r \equiv u^{-1} $, $\varphi$, and  $\vartheta = \pi/2$
are the spherical coordinates). Note
 that both the non-homogeneous wave slowness, $n^{-1} = g^{1/2}_{oo}
   \neq const$, and the non-homogeneous
 frequency,
$\hbar \omega = k_o g^{-1/2}_{oo}  \neq const$ (red shift), are responsible
 for the twofold curvature of light rays in gravitational fields.

The variations with respect to  $ u $ and
 $ \varphi $
  leads to a couple
of ray equations,
$$
(1- GM u)^4 \left [ \left({du\over d\varphi}\right )^2 +
u^2    \right] = U_o^2 = const,
$$
$$
{{d^2u}\over d\varphi^2} + u =  2U_o^2 {{GM }\over (1- GM u)^5}.
$$

A family of solutions,
 $u \equiv r^{-1}$ $ = r^{-1}_o sin \varphi $ + $2GM r^{-2}_o
(1 + cos \varphi)$ and $r_o^{-1} = U_o $, for both these equations
  might be found
in weak fields,  if one ignores all terms nonlinear in  $GMr_o^{-1}  \ll 1$.
A propagation of  light from $r(-\infty) = \infty , \varphi (-\infty)= \pi$
to  $r (+\infty)  \rightarrow  \infty , \varphi (+\infty)
 \rightarrow \varphi_{\infty} $ corresponds to the angular deflection
 $ \varphi_{\infty} = arsin [-2GM r_o^{-1}(1 + cos\varphi_\infty)]$ $\approx
  -4GMr_o^{-1}$ from the initial light direction.
This result, derived under the flat three-interval $dl$, coincides with
   the measured   deflection, $-1,66''  \pm 0.18''$ [26],
 of light rays by  the Sun  ($r_o = 6,96\times 10^5km$
and $4GM r_o^{-1} $ $=$ $1,75''$).

\bigskip \bigskip
\noindent {\large {\bf References}}

\bigskip

\noindent 1. A. Einstein. { Annalen der Physik} ${\bf 49}$, 769 (1916).

\noindent 2. R.C. Tolman  and  T.D.  Steward.
 {Phys. Rev.} ${\bf 8}$,  97 (1916).

\noindent 3. L.I. Schiff   and M.V. Barnhill.
 {Phys. Rev.} ${\bf 151}$, 1067 (1966).

\noindent 4. J.D. Jackson. { Classical electrodynamics}. New York,
John Willey and Sons, Inc. 1962.

\noindent 5. A.O. Barut.  { Electrodynamics
and Classical Theory of Fields and Particles.} New York,
The MacMillan Company. 1964.

\noindent 6.  F. Rohrlich. { Classical Charged Particles.}
Reading, Massachusetts,  Addison - Wesley Publishing Com. 1965.

\noindent 7.  D.K. Sen.  { Fields and/or Particles.}
London, Acad. Press Inc. 1968.

 \noindent 8.  L.D. Landau  and  E.M. Lifshitz.
{ The classical theory of fields.} Oxford, Pergamon. 1971.

\noindent 9.  G. Mie. { Ann. Physik.} {\bf 37},   511 (1912).

\noindent 10. M. Born  and L. Infeld.
 { Proc. R. Soc. }{\bf 144}, 425 (1934).

\noindent 11. M. - A. Tonnelat. { The principles of Electromagnetic theory
and Relativity}. Dordrecht,  D Reidel Publishing Co. 1966.

\noindent 12.  J.A. Wheeler and R.P. Feynman.
 {Rev. Mod. Phys.} {\bf 21},  425 (1949); A.D. Fokker.  {Physica}
{\bf 12}, 145 (1932).

\noindent 13. B. Felch,  J. Tate, B. Cabrera and J. T. Anderson.
 {Phys. Rev. Lett.} {\bf 62}, 845 (1989).

\noindent 14.  S.B. Felch, J. Tate and B. Cabrera.
{Phys. Rev.} B {\bf  42 },
7885 (1990).

\noindent 15. Y. Aharonov and D. Bohm.  {  Phys. Rev.} {\bf 115},
485 (1959).

\noindent 16.  M. Born.  {Ann. Phys.} {Bd 30},  1 (1909).

\noindent 17. H. Weyl.  {\it Raum - Zeit - Materie}. Berlin. 1918.

\noindent 18. J.L. Anderson.  { Principles of Relativity
Physics, p. 193}. New York - London, Academic Press. 1967.

\noindent 19. D. Hilbert. {Nachrichten von der Gesellschaft
der Wissenschaften zu Gottingen} {\bf 4},  21 (1917).

\noindent 20. G. Y. Rainich.  {Trans. Am. Math. Soc.} {\bf 27},
106 (1925).

\noindent  21.  C. Misner  and J. Wheeler.
{Ann. of Physics} {\bf 2}, 525 (1957).

\noindent 22. K. Schwarzschild.
{ Kl. Math.-Phys. Tech.}.
Berlin, Sitzber. Deut. Akad. Wiss.,  424 (1916).

\noindent 23. S. Weinberg. { Gravitation and Cosmology}.
New York, John Wiley and Sons. 1972.

\noindent 24. R.M. Wald.  { General Relativity}. Chicago,
 the University of Chicago Press. 1984.

\noindent 25. C.W. Misner, K.S. Thorne and J.A. Wheeler.
{ Gravitation}. San Francisco, Freeman. 1973.

\noindent 26. C.M. Will. { Theory and experiment in gravitational physics}.
 Cambridge, Cambridge University Press. 1981.

\noindent 27. J.V. Narlikar. { A Random Walk in Relativity
and Cosmology, p.170}.  A Halsted Press Book edited by N. Dadhich,
J. Krishnarao,
   J.V. Narlikar and C.V. Vishveshwara. 1985.;
J.V. Narlikar and T. Padmanabhan.  Foundations on Physics
{\bf 18}, 659 (1988).

\noindent 28. A. Sakharov. {Soviet Physics - Doklady } {\bf 12}, 1040
(1968);
 { Theor. Math. Phys.} {\bf 23}, 435 (1976).

\noindent 29. A.A. Logunov. {Physics - Uspekhi } {\bf 38}, 179 (1995);
A.A. Logunov and M. A. Mestvirish\-vi\-li,  { Theor. Math. Phys.}
{\bf 110}, 2 (1997).

\noindent 30. H. Yilmaz.  {Nuovo Cimento B} {\bf 107}, 941 (1992).

\noindent 31. H. Ringermacher, { Class. Quant. Gravity} {\bf 11},
 2383 (1994); J. Vargas {Found. Phys.} {\bf 21}, 379 (1991).

\noindent 32. {\it Modern Kaluza-Klein theories} eds. T. Appelquist,
A. Chodos and P. G. O. Freund. Menlo Park, Addison-Wesley. 1987;
J.M. Overduin  and P.S. Wesson.  {Phys. Reports } {\bf 283}, 303 (1997).

\noindent 33. W.A. Barker.  { US Patent } {No 5,076,971}  (1991).

\noindent 34. M.A. Tamers.  { Nature} {\bf 339}, 588 (1989).

\noindent 35. E.J. Saxl. {Nature} {\bf 203}, 136 (1964).

\noindent 36. A. Einstein. { The Meaning of Relativity}.
Princeton, Princeton University Press. 1956.
 { Appendix {\uppercase\expandafter
{\romannumeral 2} }}

\noindent 37. R. Weiss. Rev. Mod. Phys. {\bf 71}, S187 (1999).

\noindent 38. J.L. Synge.  {Relativity, the General Theory}.
Amsterdam, North-Holland Publishing Company. 1960.

\end {document}